\documentclass[aps,
prb,
reprint,
superscriptaddress]{revtex4-2}
\usepackage{graphicx}
\usepackage{dcolumn}
\usepackage{bm}
\usepackage[mathlines]{lineno}
\usepackage{color}
\usepackage{xcolor}
\usepackage{amsmath}
\usepackage{amssymb}
\usepackage{setspace}
\usepackage[normalem]{ulem}

\graphicspath{{figure2/}}

\begin{document}

\title{Griffiths-like region explains the dynamic anomaly in metallic glass-forming liquids}

\author{Lin Ma}
\affiliation{Institute of Nonequilibrium Systems, School of Systems Science, Beijing Normal University, Beijing 100875, China}
\author{Xiaodong Yang}
\affiliation{Institute of Nonequilibrium Systems, School of Systems Science, Beijing Normal University, Beijing 100875, China}
\author{Xinjia Zhou}
\affiliation{Institute of Nonequilibrium Systems, School of Systems Science, Beijing Normal University, Beijing 100875, China}
\author{Gang Sun}
\email{gangsun@bnu.edu.cn}
\affiliation{Center for Advanced Quantum Studies, School of Physics and Astronomy, Beijing Normal University, Beijing 100875, China}
\author{Zhen Wei Wu}
\email{zwwu@bnu.edu.cn}
\affiliation{Institute of Nonequilibrium Systems, School of Systems Science, Beijing Normal University, Beijing 100875, China}

\date{\today}

\begin{abstract}
Complex fluids such as water exhibits many anomalous phenomena, and research suggests these properties are closely tied to critical fluctuations near the liquid-liquid phase transition critical point (LLCP). However, whether a similar LLCP exists in metallic glass-forming liquids, which are notable for their high atomic coordination, remains an open question. Although dynamic anomalies such as the breakdown of the Stokes-Einstein (SE) relation have often been attributed to dynamic heterogeneity or structural changes, relatively few studies have analyzed these anomalies from a thermodynamic-fluctuation perspective. This gap probably stems from the challenges in detecting density-driven phase transitions in such systems. Here, we use numerical simulations to explore the thermodynamic mechanisms behind dynamic anomalies in a prototypical metallic glass-forming melt. We observe substantial thermodynamic fluctuations near a particular region, which likely corresponds to a frustration state of liquid, vapor, and glass. These fluctuations may contribute to the violation of the SE relation. Our findings offer a fresh Griffiths-like perspective on the dynamic anomalies seen in supercooled metallic liquids, and shed new light on their underlying mechanisms.
\end{abstract}

\maketitle


\section{\label{Introduction}Introduction}
In 1984, Mishima {\it et al.} identified both low-density amorphous and high-density amorphous forms of ice, reporting a first-order transition between these states~\cite{Mishima1984Melting}. Given that amorphous states typically form when particle arrangements fail to crystallize during rapid cooling, the emergence of distinct high-density and low-density amorphous states has been attributed to underlying liquid-liquid phase transitions (LLT) occurring between corresponding high-density liquids and low-density liquids~\cite{Mishima1998Decompression,Gallo2016Water,Tulk2019Absence,Fuentes2019sNature}. Analogous to conventional gas-liquid phase transitions, these LLTs are characterized as first-order transitions, with coexistence lines terminating at a liquid-liquid critical point. In recent years, the concept of a LLT has frequently been invoked to account for numerous anomalous behaviors of water and has garnered extensive scientific interest~\cite{Poole1992Phase,Sciortino1997Line,Mishima1998Decompression,Sciortino2003Physiscs,Xu2005Realtion}. Since it provides a relatively unified thermodynamic underpinning that the approach to an LLT critical point enhance structural fluctuations and correlation lengths, leading to anomalies in dynamics.

As a result, LLT phenomena have been identified across diverse systems, such as elemental substances~\cite{Deringer2021Origins, Henry2020Liquid, Monaco2003Nature} and molecular liquid mixtures~\cite{Kurita2005On,Zhu2015Possible,Murata2013General,Tanaka2004Liquid}. More specially, in 2007, Sheng et al. observed a transition from low-density amorphous to high-density amorphous states in the Ce$_{55}$Al$_{45}$ amorphous alloy system, attributing this behavior to volume collapse stemming from the delocalization of $4f$ electrons in cerium~\cite{Sheng2007Polyamorphism}. Subsequently, LLTs have been documented in various amorphous alloy melts, such as Zr-, La-, and Pd-based metallic glass-formers~\cite{Wei2013Liquid,Xu2015Evidence,Lan2017Hidden,Stolpe2016Structural}.

However, since metallic supercooled liquids generally exhibit high coordination and lack electron delocalization comparable to Ce-based alloys, observing clear first-order phase transitions driven by local density variations remains challenging. 
Many studies instead interpret LLTs in these metallic liquids as driven primarily by structural reorganizations rather than significant density changes~\cite{Wei2013Liquid,Xu2015Evidence,Tanaka2020Liquid,Tanaka2000General}. Given the high coordination of metallic supercooled liquids, clear first-order LLTs induced by substantial density fluctuations are rare. As a result, relatively few studies have investigated their dynamical anomalies from thermodynamic and fluctuation-based perspectives. For instance, the breakdown of the Stokes-Einstein (SE) relation in supercooled metallic liquids is often attributed to dynamic heterogeneity~\cite{Becker2006Fractional,Takeshi2017Identifying,Han2011Transition} or local structural transformations~\cite{Xu2009Appearance,Wu2020Revisiting}, nevertheless, interpretations based on thermodynamic fluctuations remain comparatively underexplored.

In this work, using a representative metallic glass-forming system, we aim to provide a thermodynamic perspective on anomalous dynamical behaviors in disordered systems characterized by high atomic coordination. By systematically analyzing the breakdown of the SE relation in conjunction with thermodynamic fluctuations quantified by the isobaric heat capacity $C_p$ under various pressures, we find that the locus marking the onset of SE breakdown closely tracks the extrema of $C_p$. Notably, both sets of characteristic temperatures asymptotically approach a distinct thermodynamic region, identified as the intersections of the kinetic glass-transition line and the gas–liquid spinodal line. These findings suggest that the longstanding question of whether the breakdown of the SE relation in high-coordination systems can be understood from a thermodynamic-fluctuation perspective may, at least partially, be resolved, and they naturally motivate an interpretation in terms of a Griffiths-like singularity.

From a broader perspective, our results invite comparison with the notion of a Griffiths-like regime, suggesting that the anomalous dynamics observed near the intersections of the kinetic glass-transition line and the liquid–gas spinodal can be viewed in analogy with Griffiths-like smeared singularities. In particular, near the Griffiths-like regions ($G$-region), metallic glass-forming systems may exhibit local heterogeneities characterized by fluctuations in density or structure~\cite{Wei2013Liquid,Xu2015Evidence,wu2016critical,Wu2020Revisiting,zhou2025graph}, resembling the rare-region effects of Griffiths phenomena. These spatially and temporally heterogeneous regions can facilitate unusually large thermodynamic fluctuations, dramatically influencing the local dynamics and contributing to the observed breakdown of the SE relation. In this sense, we employ the term ``Griffiths-like'' as a phenomenological description that links the enhanced thermodynamic response and the dynamic decoupling to the presence of rare, distinctly relaxing domains emerging in the vicinity of the $G$-region under negative pressure. This thermo–mechanical oriented viewpoint, rooted in negative-pressure stability and proximity to a spinodal-like instability, provides a useful interpretive lens for the observed transport anomalies in metallic glass-forming liquids.

\section{\label{Method}Model and Method}
We performed molecular dynamics (MD) simulations to investigate the structural evolution of a Cu$_{50}$Zr$_{50}$ metallic alloy as a function of temperature and pressure using the LAMMPS package~\cite{Plimpton1995Fast}. The system consists of 10,000 atoms initially distributed randomly within a cubic simulation box with periodic boundary conditions. Interatomic interactions were described using a realistic embedded-atom method (EAM) potential~\cite{Mendelev2007Using}. All MD simulations employed a time step of 2~fs. The sample was first equilibrated at $T=3500$K under zero-pressure ($P=0$~GPa) conditions using the NPT ensemble for 2,000,000 MD steps. An additional equilibration for 1,000,000 MD steps was then carried out to generate 20 independent atomic configurations. These configurations were subsequently cooled to $T=500$K to examine the system's structural transitions, as well as its dynamical and thermodynamic responses during cooling. Two distinct cooling protocols were employed: constant-volume (isodensity) cooling and constant-pressure (isobaric) cooling.

\section{\label{Results}Results}

\begin{figure}[ht]
\centering
	\includegraphics[width=\columnwidth]{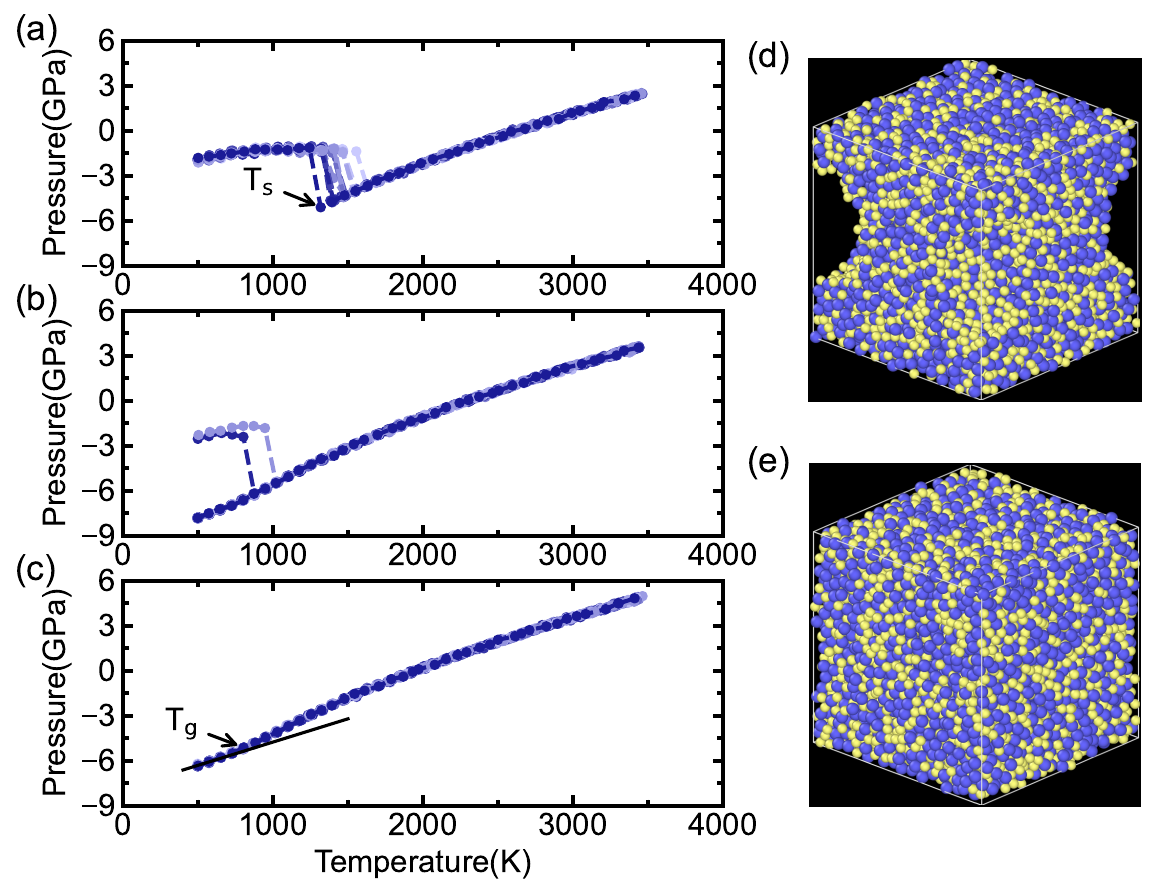}
	\caption{(a-c) Variation of pressure with temperature during isodensity cooling at different number densities: (a) $\rho = 50$, (b) $\rho = 51$, and (c) $\rho = 52$ (in units of $\text{nm}^{-3}$). Each curve represents one of the 20 independent configurations. The definition of $T_s$ and $T_g$ used in this work was also given in the panel accordingly. (d-e) Snapshots of atomic configurations at the end of the cooling process ($T=500$K): (d) $\rho = 50$, showing bubble formation; (e) $\rho = 52$, exhibiting a homogeneous glassy state. Note that here the reported $T_{\rm s}$ is the locus of spinodal points where the the homogeneous liquid approaches its mechanical stability limit.}
    \label{Fig:isodens}
\end{figure}

\begin{figure}[ht]
\centering
\includegraphics[width=\columnwidth]{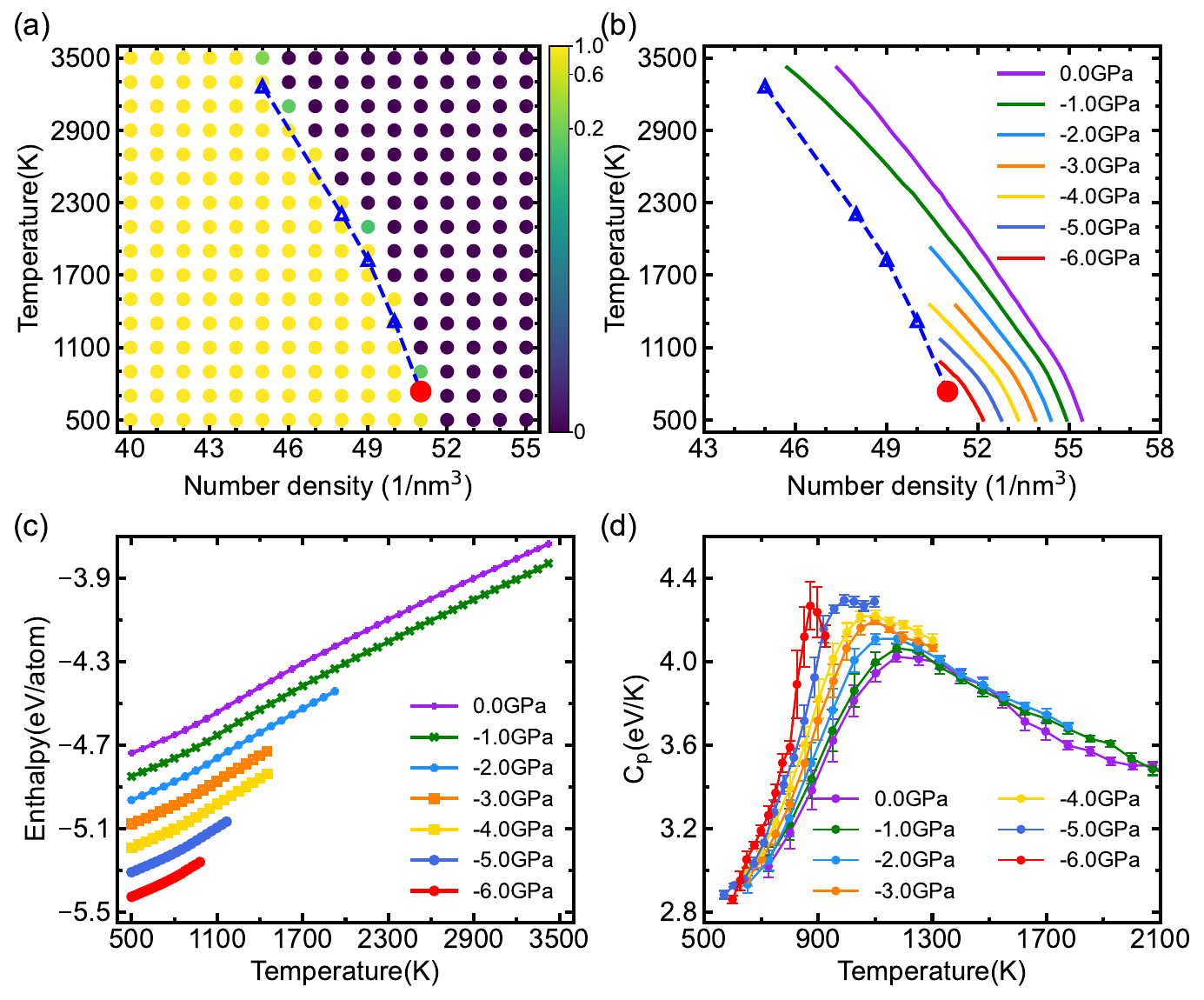}
	\caption{(a) A diagram indicating the occurrence of pressure discontinuities for 20 independent configurations prepared at various target number densities and temperatures. The color of each point denotes the probability of observing a pressure discontinuity at the corresponding state point. Blue triangles represent the characteristic temperatures ($T_s$) at which pressure jumps occur during the isodensity cooling processes, as described in detail in the main text.
    Variation of (b) number density and (c) enthalpy as a function of temperature during isobaric cooling at various pressures. (d) Corresponding temperature dependence of the isobaric heat capacity $C_p$ with error bars (using the method described in the Appendix), highlighting pronounced maxima at specific pressures.}
    \label{Fig:thermo}
\end{figure}

{\it Isodensity cooling---}
Figure~\ref{Fig:isodens} presents the variation of pressure with temperature during isodensity cooling at a rate of 0.75 K/ps for 20 independent configurations. Each curve corresponds to the cooling history of a single configuration. At a number density of $\rho = 50$ ($\text{nm}^{-3}$), all 20 configurations exhibit a discontinuous pressure jump upon cooling. From a microscopic perspective, the observed sudden change in pressure corresponds directly to bubble formation within the system. Specifically, reducing the pressure below the liquid's equilibrium vapor pressure places it into a metastable state of tension. All liquids possess measurable tensile strength, hence under sufficiently large tension the liquid spontaneously pulls apart, generating small vapor cavities (bubbles) in a process known as cavitation. Fig.~\ref{Fig:isodens}(d) illustrates a snapshot of the final atomic configurations at $T=500$K, clearly displaying these bubbles. At a slightly higher density $\rho = 51$ ($\text{nm}^{-3}$), however, only two out of the 20 configurations show a similar discontinuous pressure jumps, indicating that bubble formation becomes significantly less prevalent. For $\rho = 52$ ($\text{nm}^{-3}$), no discontinuous changes in pressure are observed at this time, during the coolings in any of the 20 configurations. Instead, the pressure changes continuously as a function of temperature, thus no bubble formation occurs, and the system exhibits a continuous liquid to glass transition. In Fig.~\ref{Fig:isodens}(c), $T_g$ is marked to denote the temperature at which there is a bend in the $P$-$T$ curve. Fig.~\ref{Fig:isodens}(e) confirms the absence of bubble formation at this number density, demonstrating that the system has translated into a ``homogeneous'' glass state. These observations highlight the sensitive dependence of bubble formation and associated pressure discontinuities on the initial number density of the configurations during the isodensity cooling. Consequently, at each density where pressure discontinuities were observed, we defined the characteristic temperature $T_s$, marked by the arrow in Fig.~\ref{Fig:isodens}(a)), as the lowest temperature among the 20 configurations at which a discontinuous pressure change occurs, marked by a blue triangle in Fig.~\ref{Fig:thermo}(a).  At the specific number density $\rho = 51\,\text{nm}^{-3}$, bubble formation occurs sporadically alongside glass formation, indicating a ``coexistence'' of liquid, bubble, and glass states. We highlight this unique {\it smeared} ``point'' with a red solid circle in Fig.~\ref{Fig:thermo}(a).

To further explore the occurrence of pressure discontinuities, specifically the liquid-gas spinodal line and coexistence region as briefly mentioned above, we independently prepared 20 configurations at targeted densities and temperatures, equilibrating each for 1,000,000 MD steps. The results are also illustrated in Fig.~\ref{Fig:thermo}(a), where the color of each dot indicates the probability of observing a pressure discontinuity at the corresponding state point. Yellow dots (probability = 1) denote states where all 20 configurations exhibit discontinuous pressure changes, indicating universal bubble formation. Purple dots (probability = 0) represent conditions where all configurations exhibit continuous pressure evolution, corresponding to uniform glass formation without bubbles. Intermediate-colored dots indicate conditions with partial or probabilistic bubble formation, reflecting cases where pressure discontinuities occur only in some configurations. Notably, the blue triangles consistently appear near the boundary separating yellow and purple regions, suggesting that the temperature at which pressure discontinuities occur is largely independent of the cooling protocol, remaining consistent whether the system is directly equilibrated at the target conditions or subjected to isodensity cooling.

{\it Isobaric cooling---}
Figure~\ref{Fig:thermo} also illustrates the evolution of the number density (panel b) and enthalpy (panel c) during the isobaric cooling process, with each curve representing the average of $20$ independent configurations at a given pressure. In panel (b) we highlight the temperatures corresponding to the pressure discontinuities previously identified in the isodensity cooling process (see Fig.~\ref{Fig:thermo}(a)) again. As bubble formation tends to occur at higher temperatures under low-density (negative pressure) conditions, the starting temperature for isobaric cooling at these lower pressures was adjusted to values below $3500\,\text{K}$. As demonstrated in Fig.~\ref{Fig:thermo}(b), decreasing the temperature under constant pressure reduces system volume, leading to an increase in number density. Correspondingly, Fig.~\ref{Fig:thermo}(c) shows a monotonic decrease in enthalpy as the system cools.

The isobaric heat capacity $C_p$, was subsequently determined from the temperature derivative of the enthalpy, as explicitly shown in Fig.~\ref{Fig:thermo}(d). During isobaric cooling, a pronounced peak in $C_p$ emerges, indicative of enhanced thermodynamic (enthalpy) fluctuations. Remarkably, as pressure decreases, the peak magnitude of $C_p$ increases, while the corresponding peak temperature systematically shifts to lower values, demonstrating asymptotic behavior toward the $G$-region (further discussed below).

\begin{figure}[htb]
\centering
\includegraphics[width=\columnwidth]{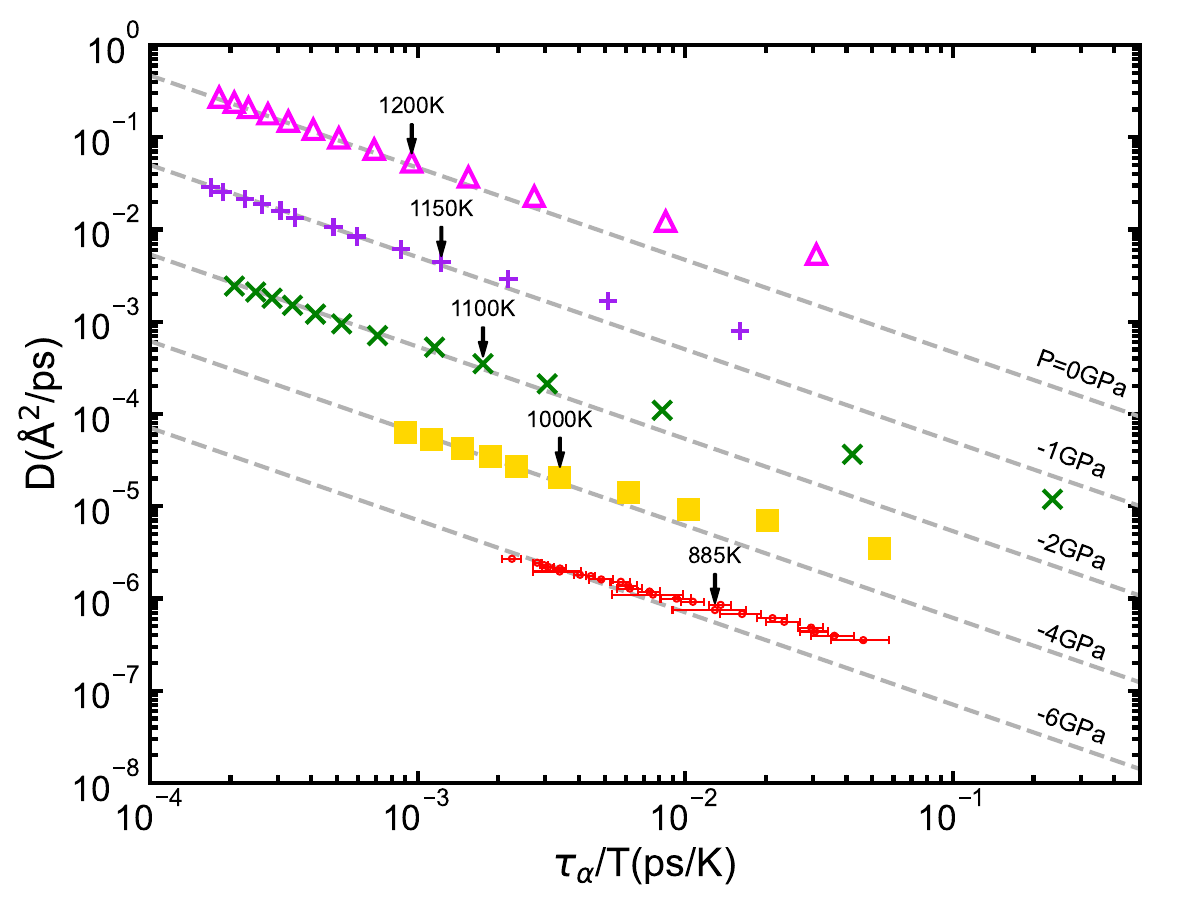}
\caption{Relationship between the diffusion coefficient $D$ and relaxation time $\tau_{\alpha}$ (scaled by $T$) during the isobaric cooling process at five different pressures. The dashed grey line represents the Stokes-Einstein relation, $D \propto (\tau_{\alpha}/T)^{-1}$. The arrows indicate the temperatures below which the SE relation breaks down. Curves for $P<0$ GPa are shifted downward by multiple factors of 0.1 for visibility. Specifically, at $P = -6\,\text{GPa}$, we include error bars to represent statistical variability arising from cavitation events.}
\label{Fig:ser}
\end{figure}

We next analyzed the dynamic properties of the system to elucidate the relationship between the Griffiths-like smeared singularity and the breakdown of the SE relation. The SE relation typically manifests as $D \sim (\tau_{\alpha}/T)^{-1}$, where the diffusion coefficient $D$ is defined by the long-time limit of the mean-squared displacement, $D = \lim \limits_{t \rightarrow \infty} \frac{1}{6t} \langle \Delta \bm{r}_j^2 (t) \rangle$ with $\Delta \bm{r}_j(t)$ denoting the displacement of particle $j$ over time $t$. The structural relaxation time $\tau_{\alpha}$ is extracted from the decay of the self-intermediate scattering function, $F_{\rm s}(q,t) = N^{-1} \langle \sum_{j=1}^{N} \exp [{\rm i}\bm{q} \cdot \Delta \bm{r}_{j}(t)] \rangle$, as the time at which $F_{\mathrm{s}}(q,t)$ reaches the value $e^{-1}$~\cite{Kob1995Testing}, and see Appendix for additional tests on the extraction of $\tau_{\alpha}$ from the {\it coherent} intermediate scattering function $F(q,t)$. Here, $q=2.8\,\text{\AA}^{-1}$ corresponds to the main peak in the static structure factor~\cite{Wu2018Stretched}, and the average $\langle \cdot \rangle$ is performed over all wave vectors with $|\bm{q}|=q$ and all initial configurations. Figure~\ref{Fig:ser} presents the relationship between diffusion coefficient $D$ and scaled relaxation time $\tau_{\alpha}/T$ under different isobaric conditions. At high temperatures, the SE relation is clearly obeyed, represented by the linear relation $D \sim (\tau_{\alpha}/T)^{-1}$. However, below a characteristic onset temperature, dependent upon the applied pressure, significant deviations from the SE relation become evident. Notably, this onset of SE breakdown coincides closely with the temperature range where the isobaric heat capacity $C_p$ exhibits its maximum, strongly suggesting a link between dynamical anomalies and enhanced thermodynamic fluctuations in our system.

\begin{figure}[htb]
\centering
\includegraphics[width=\columnwidth]{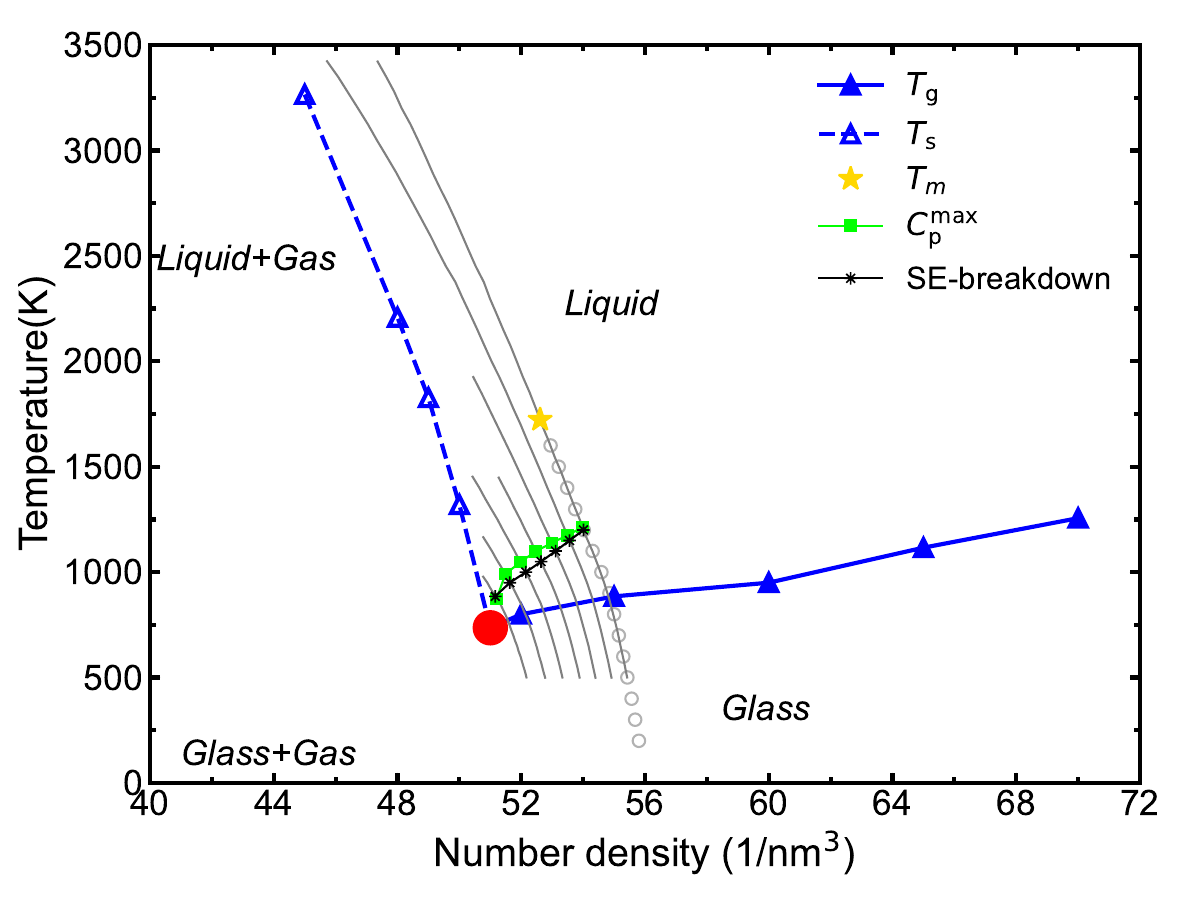}
\caption{Summary of characteristic temperatures near the Griffiths-like singularity, a smeared point representative by a red solid circle. Empty blue triangles denote the temperature $T_s$ at which pressure discontinuities occur, and solid blue triangles represent the glass-transition temperature $T_g$ identified in this study. Green squares indicate the temperature corresponding to the maximum isobaric heat capacity $C_p$, while black crosses mark the onset temperature for breakdown of the Stokes-Einstein (SE) relation. Gray lines represent the variation of number density with temperature, as previously shown in Fig.~\ref{Fig:thermo}(b). Grey circles denote data obtained from a step-by-step cooling-equilibration process, demonstrating that our conclusions are robust against minor variations in the cooling protocol. The yellow star represents the melting temperature $T_{\mathrm{m}}$ measured at zero pressure using the two-phase coexistence method.}
\label{Fig:all}
\end{figure}

To consolidate our understanding of these thermodynamic and dynamic behaviors and explore their underlying connections, we plot several characteristic temperatures simultaneously in Fig.~\ref{Fig:all}. Specifically, we illustrate the temperature $T_s$ at which pressure discontinuities occur, the glass-transition temperature $T_g$, the temperature at the maximum $C_p$, and the onset temperature for SE relation breakdown. It is noteworthy that, at different pressures, the temperature of the $C_p$ maximum closely coincides with the onset of SE breakdown. This correlation underscores a plausible thermodynamic origin for the dynamical anomalies observed in the system. Furthermore, as pressure decreases, both the $C_p$ maximum and SE breakdown temperatures shift downward and asymptotically approach the smeared $G$-point, marked by the red solid circle. This convergence implies that pronounced thermodynamic fluctuations near the $G$-region likely drive the anomalous dynamical behavior, particularly the breakdown of the SE relation.

\section{\label{Discussion and outlooks}Discussion and outlooks}

Previous studies have established that critical fluctuations near a critical point (the endpoint of a gas-liquid or liquid-liquid coexistence line) strongly enhance certain thermodynamic quantities within a specific temperature interval upon cooling. Moreover, several dynamical anomalies, such as the breakdown of the SE relation, have been closely associated with these fluctuations near critical points. Nevertheless, in metallic supercooled liquids characterized by high atomic coordination, detecting clear first-order phase transitions driven by substantial local density variations remains challenging both experimentally and computationally.

In this study, we investigate the thermodynamic mechanisms underlying dynamical anomalies in metallic glass-forming systems, taking the Cu$_{50}$Zr$_{50}$ alloy as a representative model. Our analysis reveals a distinctive thermodynamic states, defined by the intersections between the kinetic glass-transition line and the gas-liquid spinodal line. Enhanced thermodynamic fluctuations observed near this particular region closely coincide with pronounced anomalies in dynamical properties, notably the breakdown of the SE relation. As the system approaches this unique thermodynamic region, substantial concurrent anomalies emerge in both thermodynamic and dynamic behaviors.

\begin{figure*}[htb]
    \centering
    \includegraphics[width=\linewidth]{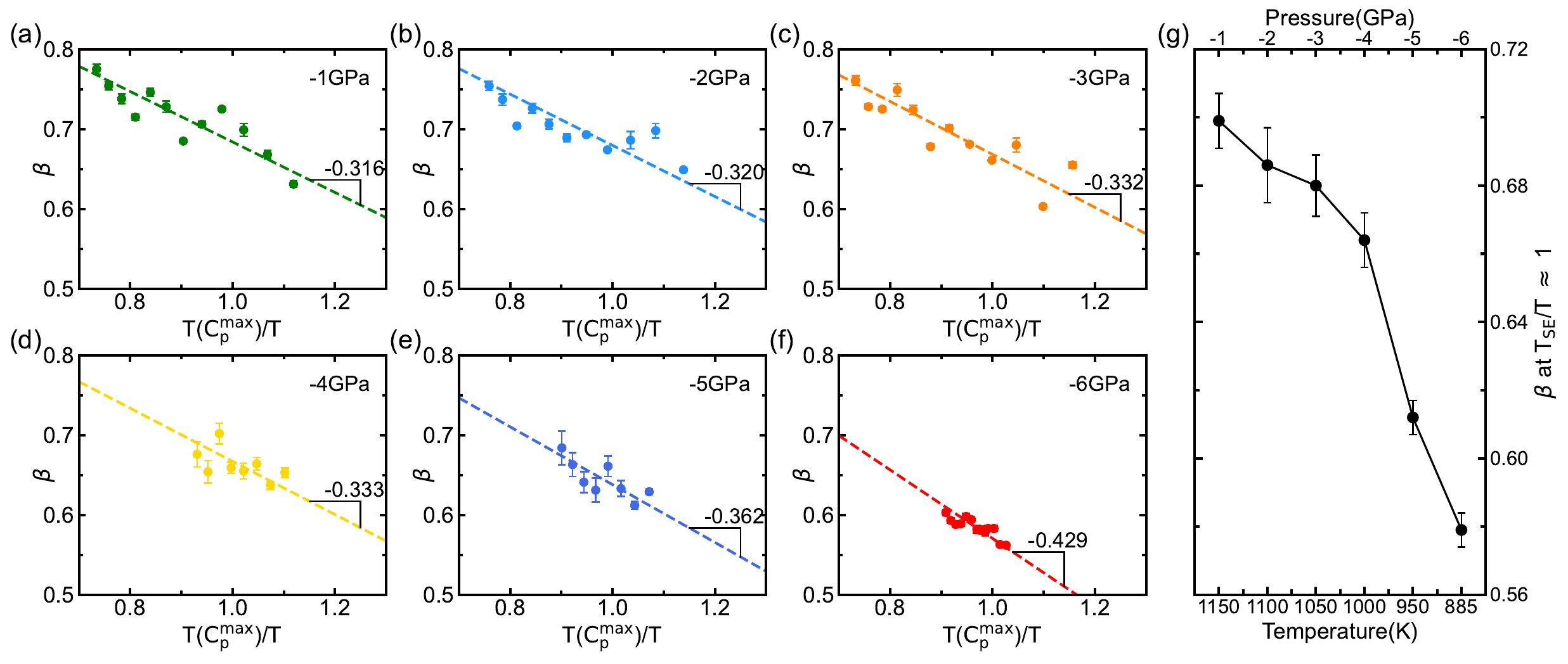}
    \caption{The Kohlrausch exponent $\beta_{\rm KWW}$ at different temperatures and pressures is obtained by fitting the corresponding $F_{\rm s}(q,t)$ with the Kohlrausch–Williams–Watts (KWW) function. Panels (a)–(f) show $\beta_{\rm KWW}(T,P)$ at pressures (a) –1~GPa, (b) –2~GPa, (c) –3~GPa, (d) –4~GPa, (e) –5~GPa, and (f) –6~GPa. Panel (g) displays $\beta_{\rm KWW}$ along the SE-violation line as the state points approach the $G$-region. Error bars in panels have been obtained from the KWW-fit.}
    \label{fig:placeholder}
\end{figure*}

The phase-diagram analysis above shows that distinct loci, i.e. the temperatures of the $C_p$ maxima and the onset of SE breakdown, converge toward a narrow region centered on the intersections of the kinetic glass-transition line with the liquid–gas spinodal. As the precise location of this intersection would slightly depend on the cooling history, resulting in a narrow ``coexistence'' region rather than a singular point, hence, as suggested in the Introduction, this behavior is naturally viewed through the lens of a Griffiths-like smeared singularity~\cite{angell1995old,Tanaka2000General}. Here we elaborate on this connection in light of our numerical results. In disordered systems with quenched or effective disorder, so-called rare regions can dominate the response, producing non-analytic yet smeared singularities in thermodynamic properties over an extended parameter range rather than at a single critical point~\cite{griffiths1969nonanalytic,vojta2006rare}. In the classical Griffiths picture, such rare regions appear as locally ``ordered islands'' embedded in a disordered ``sea'': they arise from statistically improbable fluctuations, are sufficiently large to sustain local order, relax on longer time scales, and thus exert a disproportionate influence on macroscopic observables. The geometry, size distribution, and spatial statistics of these domains determine the detailed characteristics of the Griffiths singularity. By analogy, we propose that, near the $G$-region, due to the presence of local metastable or nearly critical clusters~\cite{tanaka1999two,wu2016critical,hu2022revealing} metallic glass-forming systems may exhibit local heterogeneities characterized by fluctuations in density or structure~\cite{Wei2013Liquid,Xu2015Evidence,wu2016critical,Wu2020Revisiting,zhou2025graph}, resembling the rare-region effects of Griffiths phenomena. These spatially and temporally heterogeneous regions can facilitate unusually large thermodynamic fluctuations, dramatically influence the dynamics, and thereby contribute to the observed dynamical crossover and the decoupling from the SE relation.

Within this phenomenological Griffiths-like picture, several features of the dynamics acquire a natural interpretation. E.g., the self-intermediate scattering function exhibits an evolving stretched-exponential decay, with a drifting Kohlrausch exponent $\beta_{\rm KWW}(T,P)$ rather than saturation. As shown in Fig.~5, $\beta_{\rm KWW}$ at different temperatures and pressures is obtained by fitting $F_{\rm s}(q,t)$ with a Kohlrausch–Williams–Watts function, $F_{\rm s}(q,t) = A \exp\left[-(t/\tau)^\beta\right]$, and displays a systematic drift upon cooling. The guiding lines in Fig.~5(a)–(f) highlight that this drifting becomes more evidenced as the state point $(T,P)$ approaches the intersection region. Importantly, Fig.~5(g) shows that even along the sequence of state points on the SE-violation locus, as they approach the smeared point, $\beta_{\rm KWW}(T,P)$ continues to drift rather than saturate. This behavior is consistent with a broad crossover governed by a wide distribution of relaxation times, rather than the approach to a sharp critical point with a well-defined asymptotic exponent. The fact that these signatures become more pronounced near the $G$-region, where the $C_p$ maxima and SE-breakdown locus converge, strengthens the view that rare, anomalously dynamic domains play an increasingly important role in controlling both thermodynamic response and transport.

At the current stage, our data support a Griffiths-like description as a useful interpretive lens for the anomalous dynamics near the $G$-region, while a more detailed microscopic characterization of the underlying rare regions (e.g., their structural signatures) is a promising direction for future work. This interpretation is complementary to other established frameworks for dynamic heterogeneity, such as dynamic facilitation and cooperatively rearranging regions, which emphasize kinetic constraints and growing dynamical length scales. In contrast, our present picture is more thermo–mechanically oriented: it highlights the role of negative-pressure stability and proximity to a spinodal-like instability as a concrete thermodynamic rationale for the observed anomalies in metallic glass-forming liquids. Approaching the liquid–gas spinodal under tension constrains long-wavelength density fluctuations (stability against cavitation) while simultaneously amplifying the influence of rare regions that govern structural relaxation. We anticipate that a more detailed structural identification of these rare domains and a systematic comparison with alternative dynamical theories will provide an interesting avenue for future work and may help establish whether similar Griffiths-like regions and Griffiths-like signatures arise more generally in other metallic glass-forming compositions and, ultimately, in experiments under negative pressure.

Indeed, unlike conventional Griffiths scenarios typically discussed in the context of magnetic or spin-glass systems, where quenched disorder drives local fluctuations, the Griffiths-like scenario proposed here arises from intrinsic density and structural fluctuations that occur naturally in supercooled metallic liquids near spinodal-like instabilities. Thus, the ``Griffiths-like'' description we propose provides a useful phenomenological framework to understand the coupling between enhanced thermodynamic fluctuations and anomalous dynamics in supercooled metallic alloys, going beyond standard picture often used to interpret the dynamics of glass-forming liquids.

It is instructive to emphasize that the proposed $G$-point fundamentally differs from a conventional triple point, where three distinct phases (gas, liquid, and solid) coexist in thermodynamic equilibrium. At a typical gas-liquid-solid triple point, the solid phase corresponds to a crystalline state exhibiting spontaneous symmetry breaking. Such crystalline solids possess significant symmetry distinctions from the isotropic liquid and gas phases, precluding critical fluctuations at these points due to the clear symmetry-breaking transitions and associated latent heats. In contrast, a critical point represents the termination of a coexistence line (such as liquid-gas or liquid-liquid), at which two phases become structurally and symmetrically indistinguishable, facilitating diverging fluctuations and characteristic critical phenomena.

In conclusion, our simulations reveal a Griffiths-like region in the $(T,\rho)$ plane where thermodynamic and dynamical anomalies converge: the maxima of $C_p$ and the onset of SE breakdown track each other across pressures and approach the intersections of the kinetic glass line with the liquid–gas spinodal. This $G$-region represents a kinetic–thermodynamic band where proximity to a spinodal under negative pressure both enhances thermodynamic fluctuations and amplifies the influence of rare, particularly domains that dominate structural relaxation. This thermo–mechanically oriented, Griffiths-like perspective provides a new framework to link spinodal-related instabilities with anomalous transport in metallic glass-forming liquids, and points toward future work aimed at resolving the microscopic nature of these rare domains and testing the generality of such Griffiths-like regions in other glass-forming systems.

\begin{acknowledgments}
This work was supported by the National Natural Science Foundation of China (Grant Nos.~12474184, 52031016, and 11804027).
\end{acknowledgments}

\section*{Appendix}

\begin{figure}[ht]
\begin{center}
  \includegraphics[width=\columnwidth]{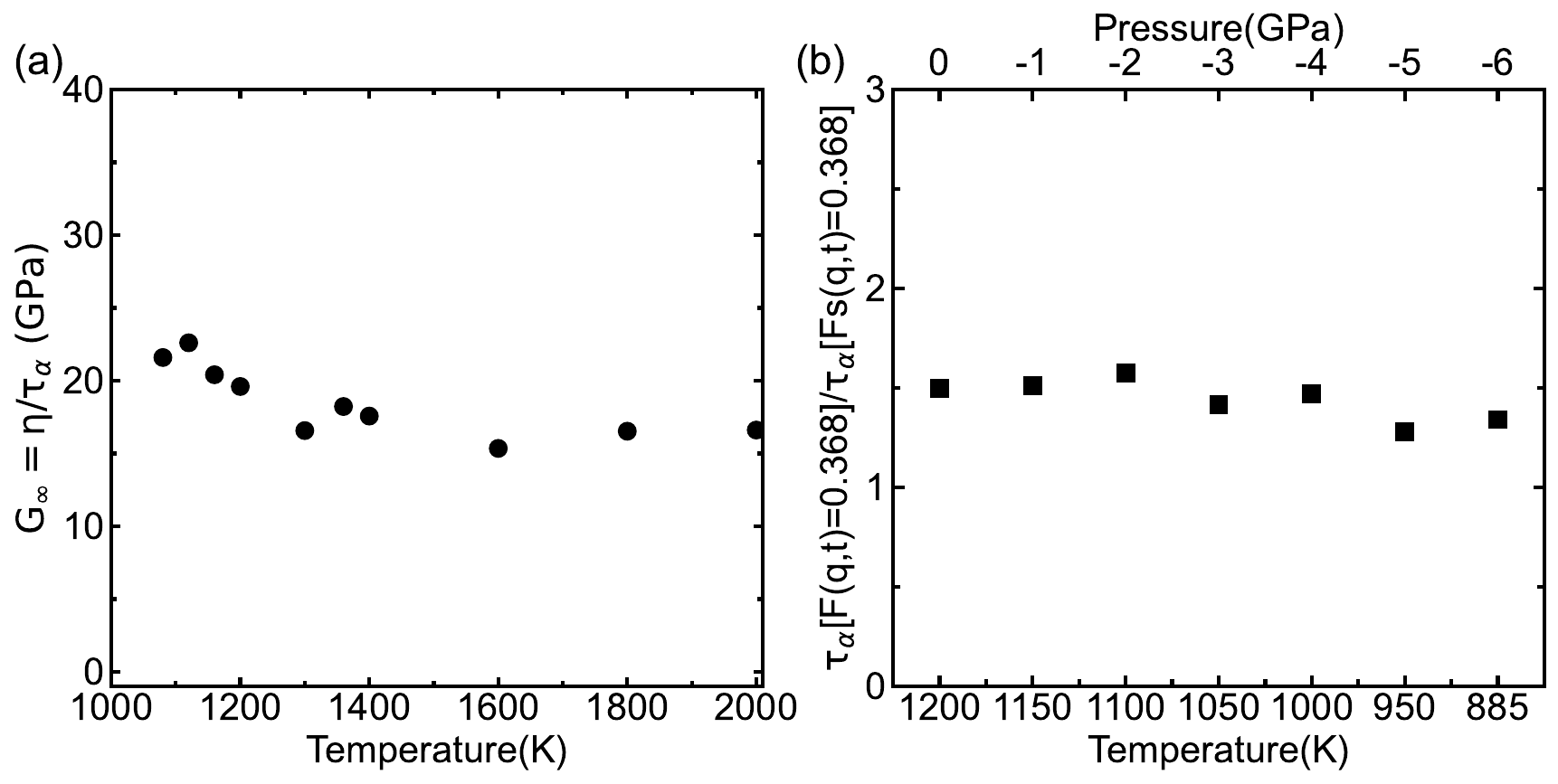}  
  \caption{
  (a) High-frequency shear modulus $G_\infty$ as a function of temperature. Data are adapted from Ref.~\cite{pan2017structural}. (b) The ratio between the structural relaxation times $\tau_\alpha$ extracted from the coherent intermediate scattering function $F(q,t)$ and the self-intermediate scattering function $F_{\rm s}(q,t)$, corresponding to the data shown in Fig.~\ref{Fig:fqt-fsqt}.
  Note that for state points of –1GPa, 1000K and –3GPa, 1400K, which lie outside the onset of SE breakdown, the corresponding ratio (which is not directly included in panel (b)) is $\sim$1.33 and $\sim$1.34, respectively.
  }
  \label{Fig:tau-eta}
\end{center}
\end{figure}

{\it Choice of $\tau_\alpha$---}
In supercooled liquids, the growth of the structural relaxation time $\tau_\alpha$ is accompanied by a concomitant increase of the shear viscosity $\eta$. This connection is commonly rationalized via the Maxwell relation in which $\eta$ and $\tau_\alpha$ are related by $\eta \simeq G_\infty \tau_\alpha$, where $G_\infty$ is the instantaneous (elastic) shear modulus~\cite{angell1995old,berthier2011theoretical}. In the supercooled regime explored here, $G_\infty(T)$ varies only moderately (see Fig.~\ref{Fig:tau-eta}(a)), especially when compared to the several orders-of-magnitude increase of both $\tau_\alpha$ and $\eta$. This justifies using $\tau_\alpha$ as a proxy for the viscosity when analyzing deviations from the Stokes–Einstein relation.

\begin{figure*}[ht]
\begin{center}
    \includegraphics[width=.9\linewidth]{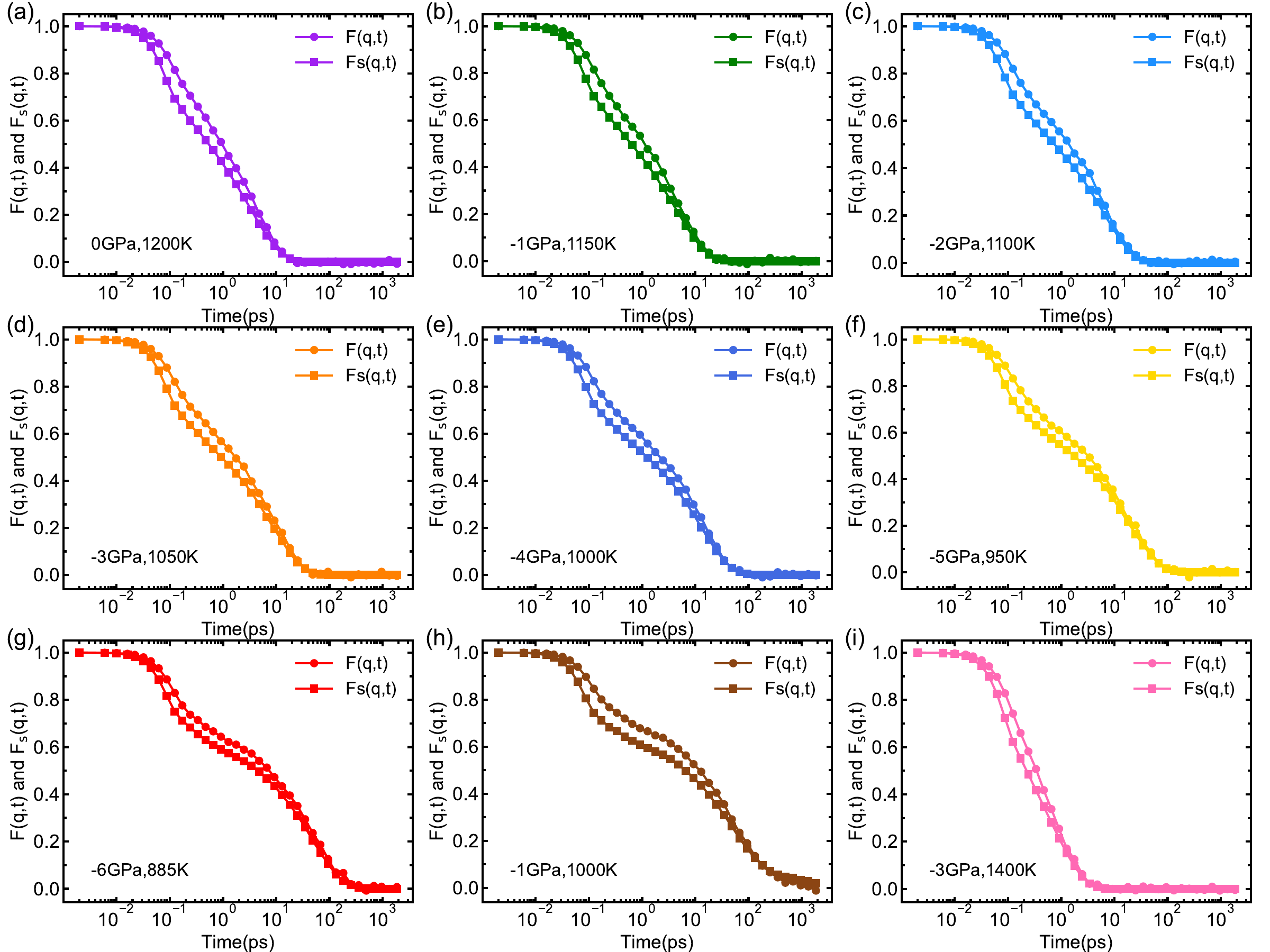}
    \caption{
    Comparison between the coherent intermediate scattering function $F(q,t)$ and the self-intermediate scattering function $F_s(q,t)$. The wave-vector is $q=2.8$\AA$^{-1}$. Panels (a)–(g) correspond to pressures ranging from 0GPa to –6GPa near the onset of Stokes–Einstein (SE) breakdown. Panels (h) and (i) present results at state points of –1GPa, 1000K and –3GPa, 1400K, respectively, which lie outside the onset of SE breakdown.
    }
    \label{Fig:fqt-fsqt}
\end{center}
\end{figure*}

We further emphasize that $\tau_\alpha$ is extracted at the first peak of the static structure factor, $q = 2.8~\text{\AA}^{-1}$, where the decay of the intermediate scattering functions is dominated by collective caging dynamics at the temperatures of interest. At this wave vector, the self-intermediate scattering function $F_{\rm s}(q,t)$ essentially probes the motion of a tagged particle within its cage and its eventual cage escape at long times. In Fig.~\ref{Fig:fqt-fsqt}, we compare the incoherent $F_{\rm s}(q,t)$ and coherent $F(q,t)$ and find that they exhibit qualitatively similar relaxation behavior at $q = 2.8~\text{\AA}^{-1}$, consistent with previous observations in other typical glass-formers~\cite{Kob1995Testing2,horbach2001relaxation}. Moreover, as shown in Fig.~\ref{Fig:tau-eta}(b), the corresponding ratio $\tau_\alpha[F(q,t)=e^{-1}]/\tau_\alpha[F_{\rm s}(q,t)=e^{-1}]$ remains close to $\sim$1.4 and, more importantly, exhibits no systematic temperature dependence over the range relevant to our study. This strongly supports the use of $\tau_\alpha$ extracted from $F_{\rm s}(q,t)$ at $q = 2.8~\text{\AA}^{-1}$ as a faithful measure of the relevant collective relaxation timescales, and consequently the crossover temperature associated with the breakdown of the SE relation is essentially unchanged whether one defines $\tau_\alpha$ from $F_{\rm s}(q,t)$ or from $F(q,t)$ at this wave-vector.

{\it $C_p$ method---}
To estimate the isobaric heat capacity $C_p(T)$ from the enthalpy--temperature data obtained during isobaric cooling, rather than employing simple finite differences, we employ a sliding-window least-squares procedure that yields a stable local estimate of the derivative $\left(\partial H/\partial T\right)_P$. For each target temperature $T_k$, we select a symmetric window of $n=2n_p+1$ neighboring data points, with $n_p=2$--$3$ points on each side. Within this window, the enthalpy is approximated by a local linear model, $H(T) \approx a_k + b_k T$. The fitted slope $b_k$ provides the local estimate of the constant-pressure heat capacity,
\begin{equation}
C_p(T_k) = b_k
= \frac{\sum_{i=k-n_p}^{k+n_p}(T_i-\overline{T})(H_i-\overline{H})}
{\sum_{i=k-n_p}^{k+n_p}(T_i-\overline{T})^2}~,
\end{equation}
where $\overline{T}$ and $\overline{H}$ denote the averages of $T_i$ and $H_i$ over the same window. The statistical uncertainty of $C_p(T_k)$ is quantified by the standard error of the fitted slope,
\begin{equation}
\mathrm{ERR} \left(b_k\right)=
\sqrt{\frac{\mathrm{RSS}}
{(n-2)\sum_{i=k-n_p}^{k+n_p}(T_i-\overline{T})^2}}~,
\end{equation}
with the residual sum of squares (RSS) defined as
\begin{equation}
\mathrm{RSS}=\sum_{i=k-n_p}^{k+n_p}\left[H_i-(a_k+b_kT_i)\right]^2~.
\end{equation}
Here, the factor $n-2$ accounts for the two fitted parameters ($a_k$ and $b_k$). This formulation directly propagates the statistical uncertainty of the local linear regression into the corresponding error bars of $C_p$.\\

\bibliography{reference}

\begin{thebibliography}{43}%
\makeatletter
\providecommand \@ifxundefined [1]{%
 \@ifx{#1\undefined}
}%
\providecommand \@ifnum [1]{%
 \ifnum #1\expandafter \@firstoftwo
 \else \expandafter \@secondoftwo
 \fi
}%
\providecommand \@ifx [1]{%
 \ifx #1\expandafter \@firstoftwo
 \else \expandafter \@secondoftwo
 \fi
}%
\providecommand \natexlab [1]{#1}%
\providecommand \enquote  [1]{``#1''}%
\providecommand \bibnamefont  [1]{#1}%
\providecommand \bibfnamefont [1]{#1}%
\providecommand \citenamefont [1]{#1}%
\providecommand \href@noop [0]{\@secondoftwo}%
\providecommand \href [0]{\begingroup \@sanitize@url \@href}%
\providecommand \@href[1]{\@@startlink{#1}\@@href}%
\providecommand \@@href[1]{\endgroup#1\@@endlink}%
\providecommand \@sanitize@url [0]{\catcode `\\12\catcode `\$12\catcode `\&12\catcode `\#12\catcode `\^12\catcode `\_12\catcode `\%12\relax}%
\providecommand \@@startlink[1]{}%
\providecommand \@@endlink[0]{}%
\providecommand \url  [0]{\begingroup\@sanitize@url \@url }%
\providecommand \@url [1]{\endgroup\@href {#1}{\urlprefix }}%
\providecommand \urlprefix  [0]{URL }%
\providecommand \Eprint [0]{\href }%
\providecommand \doibase [0]{https://doi.org/}%
\providecommand \selectlanguage [0]{\@gobble}%
\providecommand \bibinfo  [0]{\@secondoftwo}%
\providecommand \bibfield  [0]{\@secondoftwo}%
\providecommand \translation [1]{[#1]}%
\providecommand \BibitemOpen [0]{}%
\providecommand \bibitemStop [0]{}%
\providecommand \bibitemNoStop [0]{.\EOS\space}%
\providecommand \EOS [0]{\spacefactor3000\relax}%
\providecommand \BibitemShut  [1]{\csname bibitem#1\endcsname}%
\let\auto@bib@innerbib\@empty
\bibitem [{\citenamefont {Mishima}\ \emph {et~al.}(1984)\citenamefont {Mishima}, \citenamefont {Calvert},\ and\ \citenamefont {Whalley}}]{Mishima1984Melting}%
  \BibitemOpen
  \bibfield  {author} {\bibinfo {author} {\bibfnamefont {O.}~\bibnamefont {Mishima}}, \bibinfo {author} {\bibfnamefont {L.~D.}\ \bibnamefont {Calvert}},\ and\ \bibinfo {author} {\bibfnamefont {E.}~\bibnamefont {Whalley}},\ }\bibfield  {title} {\bibinfo {title} {‘{M}elting ice’ {I} at 77 {K} and 10 kbar: a new method of making amorphous solids},\ }\href {https://doi.org/10.1038/310393a0} {\bibfield  {journal} {\bibinfo  {journal} {Nature}\ }\textbf {\bibinfo {volume} {310}},\ \bibinfo {pages} {393} (\bibinfo {year} {1984})}\BibitemShut {NoStop}%
\bibitem [{\citenamefont {Mishima}\ and\ \citenamefont {Stanley}(1998)}]{Mishima1998Decompression}%
  \BibitemOpen
  \bibfield  {author} {\bibinfo {author} {\bibfnamefont {O.}~\bibnamefont {Mishima}}\ and\ \bibinfo {author} {\bibfnamefont {H.~E.}\ \bibnamefont {Stanley}},\ }\bibfield  {title} {\bibinfo {title} {Decompression-induced melting of ice iv and the liquid–liquid transition in water},\ }\href {https://doi.org/10.1038/32386} {\bibfield  {journal} {\bibinfo  {journal} {Nature}\ }\textbf {\bibinfo {volume} {392}},\ \bibinfo {pages} {164} (\bibinfo {year} {1998})}\BibitemShut {NoStop}%
\bibitem [{\citenamefont {Gallo}\ \emph {et~al.}(2016)\citenamefont {Gallo}, \citenamefont {Amann-Winkel}, \citenamefont {Angell}, \citenamefont {Anisimov}, \citenamefont {Caupin}, \citenamefont {Chakravarty}, \citenamefont {Lascaris}, \citenamefont {Loerting}, \citenamefont {Panagiotopoulos}, \citenamefont {Russo}, \citenamefont {Sellberg}, \citenamefont {Stanley}, \citenamefont {Tanaka}, \citenamefont {Vega}, \citenamefont {Xu},\ and\ \citenamefont {Pettersson}}]{Gallo2016Water}%
  \BibitemOpen
  \bibfield  {author} {\bibinfo {author} {\bibfnamefont {P.}~\bibnamefont {Gallo}}, \bibinfo {author} {\bibfnamefont {K.}~\bibnamefont {Amann-Winkel}}, \bibinfo {author} {\bibfnamefont {C.~A.}\ \bibnamefont {Angell}}, \bibinfo {author} {\bibfnamefont {M.~A.}\ \bibnamefont {Anisimov}}, \bibinfo {author} {\bibfnamefont {F.}~\bibnamefont {Caupin}}, \bibinfo {author} {\bibfnamefont {C.}~\bibnamefont {Chakravarty}}, \bibinfo {author} {\bibfnamefont {E.}~\bibnamefont {Lascaris}}, \bibinfo {author} {\bibfnamefont {T.}~\bibnamefont {Loerting}}, \bibinfo {author} {\bibfnamefont {A.~Z.}\ \bibnamefont {Panagiotopoulos}}, \bibinfo {author} {\bibfnamefont {J.}~\bibnamefont {Russo}}, \bibinfo {author} {\bibfnamefont {J.~A.}\ \bibnamefont {Sellberg}}, \bibinfo {author} {\bibfnamefont {H.~E.}\ \bibnamefont {Stanley}}, \bibinfo {author} {\bibfnamefont {H.}~\bibnamefont {Tanaka}}, \bibinfo {author} {\bibfnamefont {C.}~\bibnamefont {Vega}}, \bibinfo {author} {\bibfnamefont {L.}~\bibnamefont {Xu}},\ and\ \bibinfo {author}
  {\bibfnamefont {L.~G.~M.}\ \bibnamefont {Pettersson}},\ }\bibfield  {title} {\bibinfo {title} {Water: A tale of two liquids},\ }\href {https://doi.org/10.1021/acs.chemrev.5b00750} {\bibfield  {journal} {\bibinfo  {journal} {Chem. Rev.}\ }\textbf {\bibinfo {volume} {116}},\ \bibinfo {pages} {7463} (\bibinfo {year} {2016})}\BibitemShut {NoStop}%
\bibitem [{\citenamefont {Tulk}\ \emph {et~al.}(2019)\citenamefont {Tulk}, \citenamefont {Molaison}, \citenamefont {Makhluf}, \citenamefont {Manning},\ and\ \citenamefont {Klug}}]{Tulk2019Absence}%
  \BibitemOpen
  \bibfield  {author} {\bibinfo {author} {\bibfnamefont {C.~A.}\ \bibnamefont {Tulk}}, \bibinfo {author} {\bibfnamefont {J.~J.}\ \bibnamefont {Molaison}}, \bibinfo {author} {\bibfnamefont {A.~R.}\ \bibnamefont {Makhluf}}, \bibinfo {author} {\bibfnamefont {C.~E.}\ \bibnamefont {Manning}},\ and\ \bibinfo {author} {\bibfnamefont {D.~D.}\ \bibnamefont {Klug}},\ }\bibfield  {title} {\bibinfo {title} {Absence of amorphous forms when ice is compressed at low temperature},\ }\href {https://doi.org/10.1038/s41586-019-1204-5} {\bibfield  {journal} {\bibinfo  {journal} {Nature}\ }\textbf {\bibinfo {volume} {569}},\ \bibinfo {pages} {542} (\bibinfo {year} {2019})}\BibitemShut {NoStop}%
\bibitem [{\citenamefont {Fuentes-Landete}\ \emph {et~al.}(2019)\citenamefont {Fuentes-Landete}, \citenamefont {Plaga}, \citenamefont {Keppler}, \citenamefont {B\"ohmer},\ and\ \citenamefont {Loerting}}]{Fuentes2019sNature}%
  \BibitemOpen
  \bibfield  {author} {\bibinfo {author} {\bibfnamefont {V.}~\bibnamefont {Fuentes-Landete}}, \bibinfo {author} {\bibfnamefont {L.~J.}\ \bibnamefont {Plaga}}, \bibinfo {author} {\bibfnamefont {M.}~\bibnamefont {Keppler}}, \bibinfo {author} {\bibfnamefont {R.}~\bibnamefont {B\"ohmer}},\ and\ \bibinfo {author} {\bibfnamefont {T.}~\bibnamefont {Loerting}},\ }\bibfield  {title} {\bibinfo {title} {Nature of water's second glass transition elucidated by doping and isotope substitution experiments},\ }\href {https://doi.org/10.1103/PhysRevX.9.011015} {\bibfield  {journal} {\bibinfo  {journal} {Phys. Rev. X}\ }\textbf {\bibinfo {volume} {9}},\ \bibinfo {pages} {011015} (\bibinfo {year} {2019})}\BibitemShut {NoStop}%
\bibitem [{\citenamefont {Poole}\ \emph {et~al.}(1992)\citenamefont {Poole}, \citenamefont {Sciortino}, \citenamefont {Essmann},\ and\ \citenamefont {Stanley}}]{Poole1992Phase}%
  \BibitemOpen
  \bibfield  {author} {\bibinfo {author} {\bibfnamefont {P.~H.}\ \bibnamefont {Poole}}, \bibinfo {author} {\bibfnamefont {F.}~\bibnamefont {Sciortino}}, \bibinfo {author} {\bibfnamefont {U.}~\bibnamefont {Essmann}},\ and\ \bibinfo {author} {\bibfnamefont {H.~E.}\ \bibnamefont {Stanley}},\ }\bibfield  {title} {\bibinfo {title} {Phase behaviour of metastable water},\ }\href {https://doi.org/10.1038/360324a0} {\bibfield  {journal} {\bibinfo  {journal} {Nature}\ }\textbf {\bibinfo {volume} {360}},\ \bibinfo {pages} {324} (\bibinfo {year} {1992})}\BibitemShut {NoStop}%
\bibitem [{\citenamefont {Sciortino}\ \emph {et~al.}(1997)\citenamefont {Sciortino}, \citenamefont {Poole}, \citenamefont {Essmann},\ and\ \citenamefont {Stanley}}]{Sciortino1997Line}%
  \BibitemOpen
  \bibfield  {author} {\bibinfo {author} {\bibfnamefont {F.}~\bibnamefont {Sciortino}}, \bibinfo {author} {\bibfnamefont {P.~H.}\ \bibnamefont {Poole}}, \bibinfo {author} {\bibfnamefont {U.}~\bibnamefont {Essmann}},\ and\ \bibinfo {author} {\bibfnamefont {H.~E.}\ \bibnamefont {Stanley}},\ }\bibfield  {title} {\bibinfo {title} {Line of compressibility maxima in the phase diagram of supercooled water},\ }\href {https://doi.org/10.1103/PhysRevE.55.727} {\bibfield  {journal} {\bibinfo  {journal} {Phys. Rev. E}\ }\textbf {\bibinfo {volume} {55}},\ \bibinfo {pages} {727} (\bibinfo {year} {1997})}\BibitemShut {NoStop}%
\bibitem [{\citenamefont {Sciortino}\ \emph {et~al.}(2003)\citenamefont {Sciortino}, \citenamefont {La~Nave},\ and\ \citenamefont {Tartaglia}}]{Sciortino2003Physiscs}%
  \BibitemOpen
  \bibfield  {author} {\bibinfo {author} {\bibfnamefont {F.}~\bibnamefont {Sciortino}}, \bibinfo {author} {\bibfnamefont {E.}~\bibnamefont {La~Nave}},\ and\ \bibinfo {author} {\bibfnamefont {P.}~\bibnamefont {Tartaglia}},\ }\bibfield  {title} {\bibinfo {title} {Physics of the liquid-liquid critical point},\ }\href {https://doi.org/10.1103/PhysRevLett.91.155701} {\bibfield  {journal} {\bibinfo  {journal} {Phys. Rev. Lett.}\ }\textbf {\bibinfo {volume} {91}},\ \bibinfo {pages} {155701} (\bibinfo {year} {2003})}\BibitemShut {NoStop}%
\bibitem [{\citenamefont {Xu}\ \emph {et~al.}(2005)\citenamefont {Xu}, \citenamefont {Kumar}, \citenamefont {Buldyrev}, \citenamefont {Chen}, \citenamefont {Poole}, \citenamefont {Sciortino},\ and\ \citenamefont {Stanley}}]{Xu2005Realtion}%
  \BibitemOpen
  \bibfield  {author} {\bibinfo {author} {\bibfnamefont {L.}~\bibnamefont {Xu}}, \bibinfo {author} {\bibfnamefont {P.}~\bibnamefont {Kumar}}, \bibinfo {author} {\bibfnamefont {S.~V.}\ \bibnamefont {Buldyrev}}, \bibinfo {author} {\bibfnamefont {S.-H.}\ \bibnamefont {Chen}}, \bibinfo {author} {\bibfnamefont {P.~H.}\ \bibnamefont {Poole}}, \bibinfo {author} {\bibfnamefont {F.}~\bibnamefont {Sciortino}},\ and\ \bibinfo {author} {\bibfnamefont {H.~E.}\ \bibnamefont {Stanley}},\ }\bibfield  {title} {\bibinfo {title} {Relation between the widom line and the dynamic crossover in systems with a liquid–liquid phase transition},\ }\href {https://doi.org/10.1073/pnas.0507870102} {\bibfield  {journal} {\bibinfo  {journal} {Proc. Natl. Acad. Sci. U.S.A.}\ }\textbf {\bibinfo {volume} {102}},\ \bibinfo {pages} {16558} (\bibinfo {year} {2005})}\BibitemShut {NoStop}%
\bibitem [{\citenamefont {Deringer}\ \emph {et~al.}(2021)\citenamefont {Deringer}, \citenamefont {Bernstein}, \citenamefont {Csányi}, \citenamefont {Ben~Mahmoud}, \citenamefont {Ceriotti}, \citenamefont {Wilson}, \citenamefont {Drabold},\ and\ \citenamefont {Elliott}}]{Deringer2021Origins}%
  \BibitemOpen
  \bibfield  {author} {\bibinfo {author} {\bibfnamefont {V.~L.}\ \bibnamefont {Deringer}}, \bibinfo {author} {\bibfnamefont {N.}~\bibnamefont {Bernstein}}, \bibinfo {author} {\bibfnamefont {G.}~\bibnamefont {Csányi}}, \bibinfo {author} {\bibfnamefont {C.}~\bibnamefont {Ben~Mahmoud}}, \bibinfo {author} {\bibfnamefont {M.}~\bibnamefont {Ceriotti}}, \bibinfo {author} {\bibfnamefont {M.}~\bibnamefont {Wilson}}, \bibinfo {author} {\bibfnamefont {D.~A.}\ \bibnamefont {Drabold}},\ and\ \bibinfo {author} {\bibfnamefont {S.~R.}\ \bibnamefont {Elliott}},\ }\bibfield  {title} {\bibinfo {title} {Origins of structural and electronic transitions in disordered silicon},\ }\href {https://doi.org/10.1038/s41586-020-03072-z} {\bibfield  {journal} {\bibinfo  {journal} {Nature}\ }\textbf {\bibinfo {volume} {589}},\ \bibinfo {pages} {59} (\bibinfo {year} {2021})}\BibitemShut {NoStop}%
\bibitem [{\citenamefont {Henry}\ \emph {et~al.}(2020)\citenamefont {Henry}, \citenamefont {Mezouar}, \citenamefont {Garbarino}, \citenamefont {Sifré}, \citenamefont {Weck},\ and\ \citenamefont {Datchi}}]{Henry2020Liquid}%
  \BibitemOpen
  \bibfield  {author} {\bibinfo {author} {\bibfnamefont {L.}~\bibnamefont {Henry}}, \bibinfo {author} {\bibfnamefont {M.}~\bibnamefont {Mezouar}}, \bibinfo {author} {\bibfnamefont {G.}~\bibnamefont {Garbarino}}, \bibinfo {author} {\bibfnamefont {D.}~\bibnamefont {Sifré}}, \bibinfo {author} {\bibfnamefont {G.}~\bibnamefont {Weck}},\ and\ \bibinfo {author} {\bibfnamefont {F.}~\bibnamefont {Datchi}},\ }\bibfield  {title} {\bibinfo {title} {Liquid–liquid transition and critical point in sulfur},\ }\href {https://doi.org/10.1038/s41586-020-2593-1} {\bibfield  {journal} {\bibinfo  {journal} {Nature}\ }\textbf {\bibinfo {volume} {584}},\ \bibinfo {pages} {382} (\bibinfo {year} {2020})}\BibitemShut {NoStop}%
\bibitem [{\citenamefont {Monaco}\ \emph {et~al.}(2003)\citenamefont {Monaco}, \citenamefont {Falconi}, \citenamefont {Crichton},\ and\ \citenamefont {Mezouar}}]{Monaco2003Nature}%
  \BibitemOpen
  \bibfield  {author} {\bibinfo {author} {\bibfnamefont {G.}~\bibnamefont {Monaco}}, \bibinfo {author} {\bibfnamefont {S.}~\bibnamefont {Falconi}}, \bibinfo {author} {\bibfnamefont {W.~A.}\ \bibnamefont {Crichton}},\ and\ \bibinfo {author} {\bibfnamefont {M.}~\bibnamefont {Mezouar}},\ }\bibfield  {title} {\bibinfo {title} {Nature of the first-order phase transition in fluid phosphorus at high temperature and pressure},\ }\href {https://doi.org/10.1103/PhysRevLett.90.255701} {\bibfield  {journal} {\bibinfo  {journal} {Phys. Rev. Lett.}\ }\textbf {\bibinfo {volume} {90}},\ \bibinfo {pages} {255701} (\bibinfo {year} {2003})}\BibitemShut {NoStop}%
\bibitem [{\citenamefont {Kurita}\ and\ \citenamefont {Tanaka}(2005)}]{Kurita2005On}%
  \BibitemOpen
  \bibfield  {author} {\bibinfo {author} {\bibfnamefont {R.}~\bibnamefont {Kurita}}\ and\ \bibinfo {author} {\bibfnamefont {H.}~\bibnamefont {Tanaka}},\ }\bibfield  {title} {\bibinfo {title} {On the abundance and general nature of the liquid–liquid phase transition in molecular systems},\ }\href {https://doi.org/10.1088/0953-8984/17/27/L01} {\bibfield  {journal} {\bibinfo  {journal} {J. Phys. Condens. Matter}\ }\textbf {\bibinfo {volume} {17}},\ \bibinfo {pages} {L293} (\bibinfo {year} {2005})}\BibitemShut {NoStop}%
\bibitem [{\citenamefont {Zhu}\ \emph {et~al.}(2015)\citenamefont {Zhu}, \citenamefont {Wang}, \citenamefont {Perepezko},\ and\ \citenamefont {Yu}}]{Zhu2015Possible}%
  \BibitemOpen
  \bibfield  {author} {\bibinfo {author} {\bibfnamefont {M.}~\bibnamefont {Zhu}}, \bibinfo {author} {\bibfnamefont {J.-Q.}\ \bibnamefont {Wang}}, \bibinfo {author} {\bibfnamefont {J.~H.}\ \bibnamefont {Perepezko}},\ and\ \bibinfo {author} {\bibfnamefont {L.}~\bibnamefont {Yu}},\ }\bibfield  {title} {\bibinfo {title} {Possible existence of two amorphous phases of d-mannitol related by a first-order transition},\ }\href {https://doi.org/10.1063/1.4922543} {\bibfield  {journal} {\bibinfo  {journal} {J. Chem. Phys.}\ }\textbf {\bibinfo {volume} {142}},\ \bibinfo {pages} {244504} (\bibinfo {year} {2015})}\BibitemShut {NoStop}%
\bibitem [{\citenamefont {Murata}\ and\ \citenamefont {Tanaka}(2013)}]{Murata2013General}%
  \BibitemOpen
  \bibfield  {author} {\bibinfo {author} {\bibfnamefont {K.-i.}\ \bibnamefont {Murata}}\ and\ \bibinfo {author} {\bibfnamefont {H.}~\bibnamefont {Tanaka}},\ }\bibfield  {title} {\bibinfo {title} {General nature of liquid–liquid transition in aqueous organic solutions},\ }\href {https://doi.org/10.1038/ncomms3844} {\bibfield  {journal} {\bibinfo  {journal} {Nat. Commun.}\ }\textbf {\bibinfo {volume} {4}},\ \bibinfo {pages} {2844} (\bibinfo {year} {2013})}\BibitemShut {NoStop}%
\bibitem [{\citenamefont {Tanaka}\ \emph {et~al.}(2004)\citenamefont {Tanaka}, \citenamefont {Kurita},\ and\ \citenamefont {Mataki}}]{Tanaka2004Liquid}%
  \BibitemOpen
  \bibfield  {author} {\bibinfo {author} {\bibfnamefont {H.}~\bibnamefont {Tanaka}}, \bibinfo {author} {\bibfnamefont {R.}~\bibnamefont {Kurita}},\ and\ \bibinfo {author} {\bibfnamefont {H.}~\bibnamefont {Mataki}},\ }\bibfield  {title} {\bibinfo {title} {Liquid-liquid transition in the molecular liquid triphenyl phosphite},\ }\href {https://doi.org/10.1103/PhysRevLett.92.025701} {\bibfield  {journal} {\bibinfo  {journal} {Phys. Rev. Lett.}\ }\textbf {\bibinfo {volume} {92}},\ \bibinfo {pages} {025701} (\bibinfo {year} {2004})}\BibitemShut {NoStop}%
\bibitem [{\citenamefont {Sheng}\ \emph {et~al.}(2007)\citenamefont {Sheng}, \citenamefont {Liu}, \citenamefont {Cheng}, \citenamefont {Wen}, \citenamefont {Lee}, \citenamefont {Luo}, \citenamefont {Shastri},\ and\ \citenamefont {Ma}}]{Sheng2007Polyamorphism}%
  \BibitemOpen
  \bibfield  {author} {\bibinfo {author} {\bibfnamefont {H.~W.}\ \bibnamefont {Sheng}}, \bibinfo {author} {\bibfnamefont {H.~Z.}\ \bibnamefont {Liu}}, \bibinfo {author} {\bibfnamefont {Y.~Q.}\ \bibnamefont {Cheng}}, \bibinfo {author} {\bibfnamefont {J.}~\bibnamefont {Wen}}, \bibinfo {author} {\bibfnamefont {P.~L.}\ \bibnamefont {Lee}}, \bibinfo {author} {\bibfnamefont {W.~K.}\ \bibnamefont {Luo}}, \bibinfo {author} {\bibfnamefont {S.~D.}\ \bibnamefont {Shastri}},\ and\ \bibinfo {author} {\bibfnamefont {E.}~\bibnamefont {Ma}},\ }\bibfield  {title} {\bibinfo {title} {Polyamorphism in a metallic glass},\ }\href {https://doi.org/10.1038/nmat1839} {\bibfield  {journal} {\bibinfo  {journal} {Nat. Mater.}\ }\textbf {\bibinfo {volume} {6}},\ \bibinfo {pages} {192} (\bibinfo {year} {2007})}\BibitemShut {NoStop}%
\bibitem [{\citenamefont {Wei}\ \emph {et~al.}(2013)\citenamefont {Wei}, \citenamefont {Yang}, \citenamefont {Bednarcik}, \citenamefont {Kaban}, \citenamefont {Shuleshova}, \citenamefont {Meyer},\ and\ \citenamefont {Busch}}]{Wei2013Liquid}%
  \BibitemOpen
  \bibfield  {author} {\bibinfo {author} {\bibfnamefont {S.}~\bibnamefont {Wei}}, \bibinfo {author} {\bibfnamefont {F.}~\bibnamefont {Yang}}, \bibinfo {author} {\bibfnamefont {J.}~\bibnamefont {Bednarcik}}, \bibinfo {author} {\bibfnamefont {I.}~\bibnamefont {Kaban}}, \bibinfo {author} {\bibfnamefont {O.}~\bibnamefont {Shuleshova}}, \bibinfo {author} {\bibfnamefont {A.}~\bibnamefont {Meyer}},\ and\ \bibinfo {author} {\bibfnamefont {R.}~\bibnamefont {Busch}},\ }\bibfield  {title} {\bibinfo {title} {Liquid–liquid transition in a strong bulk metallic glass-forming liquid},\ }\href {https://doi.org/10.1038/ncomms3083} {\bibfield  {journal} {\bibinfo  {journal} {Nat. Commun.}\ }\textbf {\bibinfo {volume} {4}},\ \bibinfo {pages} {2083} (\bibinfo {year} {2013})}\BibitemShut {NoStop}%
\bibitem [{\citenamefont {Xu}\ \emph {et~al.}(2015)\citenamefont {Xu}, \citenamefont {Sandor}, \citenamefont {Yu}, \citenamefont {Ke}, \citenamefont {Zhang}, \citenamefont {Li}, \citenamefont {Wang}, \citenamefont {Liu},\ and\ \citenamefont {Wu}}]{Xu2015Evidence}%
  \BibitemOpen
  \bibfield  {author} {\bibinfo {author} {\bibfnamefont {W.}~\bibnamefont {Xu}}, \bibinfo {author} {\bibfnamefont {M.~T.}\ \bibnamefont {Sandor}}, \bibinfo {author} {\bibfnamefont {Y.}~\bibnamefont {Yu}}, \bibinfo {author} {\bibfnamefont {H.-B.}\ \bibnamefont {Ke}}, \bibinfo {author} {\bibfnamefont {H.-P.}\ \bibnamefont {Zhang}}, \bibinfo {author} {\bibfnamefont {M.-Z.}\ \bibnamefont {Li}}, \bibinfo {author} {\bibfnamefont {W.-H.}\ \bibnamefont {Wang}}, \bibinfo {author} {\bibfnamefont {L.}~\bibnamefont {Liu}},\ and\ \bibinfo {author} {\bibfnamefont {Y.}~\bibnamefont {Wu}},\ }\bibfield  {title} {\bibinfo {title} {Evidence of liquid–liquid transition in glass-forming {La}$_{50}${Al}$_{35}${Ni}$_{15}$ melt above liquidus temperature},\ }\href {https://doi.org/10.1038/ncomms8696} {\bibfield  {journal} {\bibinfo  {journal} {Nat. Commun.}\ }\textbf {\bibinfo {volume} {6}},\ \bibinfo {pages} {7696} (\bibinfo {year} {2015})}\BibitemShut {NoStop}%
\bibitem [{\citenamefont {Lan}\ \emph {et~al.}(2017)\citenamefont {Lan}, \citenamefont {Ren}, \citenamefont {Wei}, \citenamefont {Wang}, \citenamefont {Gilbert}, \citenamefont {Shibayama}, \citenamefont {Watanabe}, \citenamefont {Ohnuma},\ and\ \citenamefont {Wang}}]{Lan2017Hidden}%
  \BibitemOpen
  \bibfield  {author} {\bibinfo {author} {\bibfnamefont {S.}~\bibnamefont {Lan}}, \bibinfo {author} {\bibfnamefont {Y.}~\bibnamefont {Ren}}, \bibinfo {author} {\bibfnamefont {X.~Y.}\ \bibnamefont {Wei}}, \bibinfo {author} {\bibfnamefont {B.}~\bibnamefont {Wang}}, \bibinfo {author} {\bibfnamefont {E.~P.}\ \bibnamefont {Gilbert}}, \bibinfo {author} {\bibfnamefont {T.}~\bibnamefont {Shibayama}}, \bibinfo {author} {\bibfnamefont {S.}~\bibnamefont {Watanabe}}, \bibinfo {author} {\bibfnamefont {M.}~\bibnamefont {Ohnuma}},\ and\ \bibinfo {author} {\bibfnamefont {X.~L.}\ \bibnamefont {Wang}},\ }\bibfield  {title} {\bibinfo {title} {Hidden amorphous phase and reentrant supercooled liquid in {Pd}-{Ni}-{P} metallic glasses},\ }\href {https://doi.org/10.1038/ncomms14679} {\bibfield  {journal} {\bibinfo  {journal} {Nat. Commun.}\ }\textbf {\bibinfo {volume} {8}},\ \bibinfo {pages} {14679} (\bibinfo {year} {2017})}\BibitemShut {NoStop}%
\bibitem [{\citenamefont {Stolpe}\ \emph {et~al.}(2016)\citenamefont {Stolpe}, \citenamefont {Jonas}, \citenamefont {Wei}, \citenamefont {Evenson}, \citenamefont {Hembree}, \citenamefont {Yang}, \citenamefont {Meyer},\ and\ \citenamefont {Busch}}]{Stolpe2016Structural}%
  \BibitemOpen
  \bibfield  {author} {\bibinfo {author} {\bibfnamefont {M.}~\bibnamefont {Stolpe}}, \bibinfo {author} {\bibfnamefont {I.}~\bibnamefont {Jonas}}, \bibinfo {author} {\bibfnamefont {S.}~\bibnamefont {Wei}}, \bibinfo {author} {\bibfnamefont {Z.}~\bibnamefont {Evenson}}, \bibinfo {author} {\bibfnamefont {W.}~\bibnamefont {Hembree}}, \bibinfo {author} {\bibfnamefont {F.}~\bibnamefont {Yang}}, \bibinfo {author} {\bibfnamefont {A.}~\bibnamefont {Meyer}},\ and\ \bibinfo {author} {\bibfnamefont {R.}~\bibnamefont {Busch}},\ }\bibfield  {title} {\bibinfo {title} {Structural changes during a liquid-liquid transition in the deeply undercooled $\mathrm{Z}{\mathrm{r}}_{58.5}\mathrm{C}{\mathrm{u}}_{15.6}\mathrm{N}{\mathrm{i}}_{12.8}\mathrm{A}{\mathrm{l}}_{10.3}\mathrm{N}{\mathrm{b}}_{2.8}$ bulk metallic glass forming melt},\ }\href {https://doi.org/10.1103/PhysRevB.93.014201} {\bibfield  {journal} {\bibinfo  {journal} {Phys. Rev. B}\ }\textbf {\bibinfo {volume} {93}},\ \bibinfo {pages} {014201} (\bibinfo {year}
  {2016})}\BibitemShut {NoStop}%
\bibitem [{\citenamefont {Tanaka}(2020)}]{Tanaka2020Liquid}%
  \BibitemOpen
  \bibfield  {author} {\bibinfo {author} {\bibfnamefont {H.}~\bibnamefont {Tanaka}},\ }\bibfield  {title} {\bibinfo {title} {Liquid–liquid transition and polyamorphism},\ }\href {https://doi.org/10.1063/5.0021045} {\bibfield  {journal} {\bibinfo  {journal} {J. Chem. Phys}\ }\textbf {\bibinfo {volume} {153}},\ \bibinfo {pages} {130901} (\bibinfo {year} {2020})}\BibitemShut {NoStop}%
\bibitem [{\citenamefont {Tanaka}(2000)}]{Tanaka2000General}%
  \BibitemOpen
  \bibfield  {author} {\bibinfo {author} {\bibfnamefont {H.}~\bibnamefont {Tanaka}},\ }\bibfield  {title} {\bibinfo {title} {General view of a liquid-liquid phase transition},\ }\href {https://doi.org/10.1103/PhysRevE.62.6968} {\bibfield  {journal} {\bibinfo  {journal} {Phys. Rev. E}\ }\textbf {\bibinfo {volume} {62}},\ \bibinfo {pages} {6968} (\bibinfo {year} {2000})}\BibitemShut {NoStop}%
\bibitem [{\citenamefont {Becker}\ \emph {et~al.}(2006)\citenamefont {Becker}, \citenamefont {Poole},\ and\ \citenamefont {Starr}}]{Becker2006Fractional}%
  \BibitemOpen
  \bibfield  {author} {\bibinfo {author} {\bibfnamefont {S.~R.}\ \bibnamefont {Becker}}, \bibinfo {author} {\bibfnamefont {P.~H.}\ \bibnamefont {Poole}},\ and\ \bibinfo {author} {\bibfnamefont {F.~W.}\ \bibnamefont {Starr}},\ }\bibfield  {title} {\bibinfo {title} {Fractional stokes-einstein and debye-stokes-einstein relations in a network-forming liquid},\ }\href {https://doi.org/10.1103/PhysRevLett.97.055901} {\bibfield  {journal} {\bibinfo  {journal} {Phys. Rev. Lett.}\ }\textbf {\bibinfo {volume} {97}},\ \bibinfo {pages} {055901} (\bibinfo {year} {2006})}\BibitemShut {NoStop}%
\bibitem [{\citenamefont {Kawasaki}\ and\ \citenamefont {Kim}(2017)}]{Takeshi2017Identifying}%
  \BibitemOpen
  \bibfield  {author} {\bibinfo {author} {\bibfnamefont {T.}~\bibnamefont {Kawasaki}}\ and\ \bibinfo {author} {\bibfnamefont {K.}~\bibnamefont {Kim}},\ }\bibfield  {title} {\bibinfo {title} {Identifying time scales for violation/preservation of stokes-einstein relation in supercooled water},\ }\href {https://doi.org/10.1126/sciadv.1700399} {\bibfield  {journal} {\bibinfo  {journal} {Sci. Adv.}\ }\textbf {\bibinfo {volume} {3}},\ \bibinfo {pages} {e1700399} (\bibinfo {year} {2017})}\BibitemShut {NoStop}%
\bibitem [{\citenamefont {Han}\ and\ \citenamefont {Schober}(2011)}]{Han2011Transition}%
  \BibitemOpen
  \bibfield  {author} {\bibinfo {author} {\bibfnamefont {X.~J.}\ \bibnamefont {Han}}\ and\ \bibinfo {author} {\bibfnamefont {H.~R.}\ \bibnamefont {Schober}},\ }\bibfield  {title} {\bibinfo {title} {Transport properties and stokes-einstein relation in a computer-simulated glass-forming {Cu}$_{33.3}${Zr}$_{66.7}$ melt},\ }\href {https://doi.org/10.1103/PhysRevB.83.224201} {\bibfield  {journal} {\bibinfo  {journal} {Phys. Rev. B}\ }\textbf {\bibinfo {volume} {83}},\ \bibinfo {pages} {224201} (\bibinfo {year} {2011})}\BibitemShut {NoStop}%
\bibitem [{\citenamefont {Xu}\ \emph {et~al.}(2009)\citenamefont {Xu}, \citenamefont {Mallamace}, \citenamefont {Yan}, \citenamefont {Starr}, \citenamefont {Buldyrev},\ and\ \citenamefont {Eugene~Stanley}}]{Xu2009Appearance}%
  \BibitemOpen
  \bibfield  {author} {\bibinfo {author} {\bibfnamefont {L.}~\bibnamefont {Xu}}, \bibinfo {author} {\bibfnamefont {F.}~\bibnamefont {Mallamace}}, \bibinfo {author} {\bibfnamefont {Z.}~\bibnamefont {Yan}}, \bibinfo {author} {\bibfnamefont {F.~W.}\ \bibnamefont {Starr}}, \bibinfo {author} {\bibfnamefont {S.~V.}\ \bibnamefont {Buldyrev}},\ and\ \bibinfo {author} {\bibfnamefont {H.}~\bibnamefont {Eugene~Stanley}},\ }\bibfield  {title} {\bibinfo {title} {Appearance of a fractional stokes–einstein relation in water and a structural interpretation of its onset},\ }\href {https://doi.org/10.1038/nphys1328} {\bibfield  {journal} {\bibinfo  {journal} {Nat. Phys.}\ }\textbf {\bibinfo {volume} {5}},\ \bibinfo {pages} {565} (\bibinfo {year} {2009})}\BibitemShut {NoStop}%
\bibitem [{\citenamefont {Wu}\ and\ \citenamefont {Li}(2020)}]{Wu2020Revisiting}%
  \BibitemOpen
  \bibfield  {author} {\bibinfo {author} {\bibfnamefont {Z.~W.}\ \bibnamefont {Wu}}\ and\ \bibinfo {author} {\bibfnamefont {R.~Z.}\ \bibnamefont {Li}},\ }\bibfield  {title} {\bibinfo {title} {Revisiting the breakdown of stokes-einstein relation in glass-forming liquids with machine learning},\ }\href {https://doi.org/10.1007/s11433-020-1539-4} {\bibfield  {journal} {\bibinfo  {journal} {Sci. China Phys. Mech. Astron.}\ }\textbf {\bibinfo {volume} {63}},\ \bibinfo {pages} {276111} (\bibinfo {year} {2020})}\BibitemShut {NoStop}%
\bibitem [{\citenamefont {Wu}\ \emph {et~al.}(2016)\citenamefont {Wu}, \citenamefont {Li}, \citenamefont {Huo}, \citenamefont {Li}, \citenamefont {Wang},\ and\ \citenamefont {Liu}}]{wu2016critical}%
  \BibitemOpen
  \bibfield  {author} {\bibinfo {author} {\bibfnamefont {Z.~W.}\ \bibnamefont {Wu}}, \bibinfo {author} {\bibfnamefont {F.~X.}\ \bibnamefont {Li}}, \bibinfo {author} {\bibfnamefont {C.~W.}\ \bibnamefont {Huo}}, \bibinfo {author} {\bibfnamefont {M.~Z.}\ \bibnamefont {Li}}, \bibinfo {author} {\bibfnamefont {W.~H.}\ \bibnamefont {Wang}},\ and\ \bibinfo {author} {\bibfnamefont {K.~X.}\ \bibnamefont {Liu}},\ }\bibfield  {title} {\bibinfo {title} {Critical scaling of icosahedral medium-range order in cuzr metallic glass-forming liquids},\ }\href@noop {} {\bibfield  {journal} {\bibinfo  {journal} {Sci. Rep.}\ }\textbf {\bibinfo {volume} {6}},\ \bibinfo {pages} {35967} (\bibinfo {year} {2016})}\BibitemShut {NoStop}%
\bibitem [{\citenamefont {Zhou}\ \emph {et~al.}(2025)\citenamefont {Zhou}, \citenamefont {Yang}, \citenamefont {Yang}, \citenamefont {Ma},\ and\ \citenamefont {Wu}}]{zhou2025graph}%
  \BibitemOpen
  \bibfield  {author} {\bibinfo {author} {\bibfnamefont {X.-J.}\ \bibnamefont {Zhou}}, \bibinfo {author} {\bibfnamefont {F.}~\bibnamefont {Yang}}, \bibinfo {author} {\bibfnamefont {X.-D.}\ \bibnamefont {Yang}}, \bibinfo {author} {\bibfnamefont {L.}~\bibnamefont {Ma}},\ and\ \bibinfo {author} {\bibfnamefont {Z.-W.}\ \bibnamefont {Wu}},\ }\bibfield  {title} {\bibinfo {title} {Graph-dynamics correspondence in metallic glass-forming liquids},\ }\href@noop {} {\bibfield  {journal} {\bibinfo  {journal} {Commun. Theor. Phys.}\ }\textbf {\bibinfo {volume} {77}},\ \bibinfo {pages} {097601} (\bibinfo {year} {2025})}\BibitemShut {NoStop}%
\bibitem [{\citenamefont {Plimpton}(1995)}]{Plimpton1995Fast}%
  \BibitemOpen
  \bibfield  {author} {\bibinfo {author} {\bibfnamefont {S.}~\bibnamefont {Plimpton}},\ }\bibfield  {title} {\bibinfo {title} {Fast parallel algorithms for short-range molecular dynamics},\ }\href {https://doi.org/https://doi.org/10.1006/jcph.1995.1039} {\bibfield  {journal} {\bibinfo  {journal} {J. Comput. Phys.}\ }\textbf {\bibinfo {volume} {117}},\ \bibinfo {pages} {1} (\bibinfo {year} {1995})}\BibitemShut {NoStop}%
\bibitem [{\citenamefont {Mendelev}\ \emph {et~al.}(2007)\citenamefont {Mendelev}, \citenamefont {Sordelet},\ and\ \citenamefont {Kramer}}]{Mendelev2007Using}%
  \BibitemOpen
  \bibfield  {author} {\bibinfo {author} {\bibfnamefont {M.~I.}\ \bibnamefont {Mendelev}}, \bibinfo {author} {\bibfnamefont {D.~J.}\ \bibnamefont {Sordelet}},\ and\ \bibinfo {author} {\bibfnamefont {M.~J.}\ \bibnamefont {Kramer}},\ }\bibfield  {title} {\bibinfo {title} {Using atomistic computer simulations to analyze x-ray diffraction data from metallic glasses},\ }\href {https://doi.org/10.1063/1.2769157} {\bibfield  {journal} {\bibinfo  {journal} {J. Appl. Phys.}\ }\textbf {\bibinfo {volume} {102}},\ \bibinfo {pages} {043501} (\bibinfo {year} {2007})}\BibitemShut {NoStop}%
\bibitem [{\citenamefont {Kob}\ and\ \citenamefont {Andersen}(1995{\natexlab{a}})}]{Kob1995Testing}%
  \BibitemOpen
  \bibfield  {author} {\bibinfo {author} {\bibfnamefont {W.}~\bibnamefont {Kob}}\ and\ \bibinfo {author} {\bibfnamefont {H.~C.}\ \bibnamefont {Andersen}},\ }\bibfield  {title} {\bibinfo {title} {Testing mode-coupling theory for a supercooled binary lennard-jones mixture {I}: The van hove correlation function},\ }\href {https://doi.org/10.1103/PhysRevE.51.4626} {\bibfield  {journal} {\bibinfo  {journal} {Phys. Rev. E}\ }\textbf {\bibinfo {volume} {51}},\ \bibinfo {pages} {4626} (\bibinfo {year} {1995}{\natexlab{a}})}\BibitemShut {NoStop}%
\bibitem [{\citenamefont {Wu}\ \emph {et~al.}(2018)\citenamefont {Wu}, \citenamefont {Kob}, \citenamefont {Wang},\ and\ \citenamefont {Xu}}]{Wu2018Stretched}%
  \BibitemOpen
  \bibfield  {author} {\bibinfo {author} {\bibfnamefont {Z.~W.}\ \bibnamefont {Wu}}, \bibinfo {author} {\bibfnamefont {W.}~\bibnamefont {Kob}}, \bibinfo {author} {\bibfnamefont {W.-H.}\ \bibnamefont {Wang}},\ and\ \bibinfo {author} {\bibfnamefont {L.}~\bibnamefont {Xu}},\ }\bibfield  {title} {\bibinfo {title} {Stretched and compressed exponentials in the relaxation dynamics of a metallic glass-forming melt},\ }\href {https://doi.org/10.1038/s41467-018-07759-w} {\bibfield  {journal} {\bibinfo  {journal} {Nat. Commun.}\ }\textbf {\bibinfo {volume} {9}},\ \bibinfo {pages} {5334} (\bibinfo {year} {2018})}\BibitemShut {NoStop}%
\bibitem [{\citenamefont {Angell}(1995)}]{angell1995old}%
  \BibitemOpen
  \bibfield  {author} {\bibinfo {author} {\bibfnamefont {C.}~\bibnamefont {Angell}},\ }\bibfield  {title} {\bibinfo {title} {The old problems of glass and the glass transition, and the many new twists.},\ }\href@noop {} {\bibfield  {journal} {\bibinfo  {journal} {Proc. Natl. Acad. Sci.}\ }\textbf {\bibinfo {volume} {92}},\ \bibinfo {pages} {6675} (\bibinfo {year} {1995})}\BibitemShut {NoStop}%
\bibitem [{\citenamefont {Griffiths}(1969)}]{griffiths1969nonanalytic}%
  \BibitemOpen
  \bibfield  {author} {\bibinfo {author} {\bibfnamefont {R.~B.}\ \bibnamefont {Griffiths}},\ }\bibfield  {title} {\bibinfo {title} {Nonanalytic behavior above the critical point in a random ising ferromagnet},\ }\href@noop {} {\bibfield  {journal} {\bibinfo  {journal} {Phys. Rev. Lett.}\ }\textbf {\bibinfo {volume} {23}},\ \bibinfo {pages} {17} (\bibinfo {year} {1969})}\BibitemShut {NoStop}%
\bibitem [{\citenamefont {Vojta}(2006)}]{vojta2006rare}%
  \BibitemOpen
  \bibfield  {author} {\bibinfo {author} {\bibfnamefont {T.}~\bibnamefont {Vojta}},\ }\bibfield  {title} {\bibinfo {title} {Rare region effects at classical, quantum and nonequilibrium phase transitions},\ }\href@noop {} {\bibfield  {journal} {\bibinfo  {journal} {J. Phys. A: Math. Gen.}\ }\textbf {\bibinfo {volume} {39}},\ \bibinfo {pages} {R143} (\bibinfo {year} {2006})}\BibitemShut {NoStop}%
\bibitem [{\citenamefont {Tanaka}(1999)}]{tanaka1999two}%
  \BibitemOpen
  \bibfield  {author} {\bibinfo {author} {\bibfnamefont {H.}~\bibnamefont {Tanaka}},\ }\bibfield  {title} {\bibinfo {title} {Two-order-parameter description of liquids. i. a general model of glass transition covering its strong to fragile limit},\ }\href@noop {} {\bibfield  {journal} {\bibinfo  {journal} {J. Chem. Phys}\ }\textbf {\bibinfo {volume} {111}},\ \bibinfo {pages} {3163} (\bibinfo {year} {1999})}\BibitemShut {NoStop}%
\bibitem [{\citenamefont {Hu}\ and\ \citenamefont {Tanaka}(2022)}]{hu2022revealing}%
  \BibitemOpen
  \bibfield  {author} {\bibinfo {author} {\bibfnamefont {Y.-C.}\ \bibnamefont {Hu}}\ and\ \bibinfo {author} {\bibfnamefont {H.}~\bibnamefont {Tanaka}},\ }\bibfield  {title} {\bibinfo {title} {Revealing the role of liquid preordering in crystallisation of supercooled liquids},\ }\href@noop {} {\bibfield  {journal} {\bibinfo  {journal} {Nat. Commun.}\ }\textbf {\bibinfo {volume} {13}},\ \bibinfo {pages} {4519} (\bibinfo {year} {2022})}\BibitemShut {NoStop}%
\bibitem [{\citenamefont {Berthier}\ and\ \citenamefont {Biroli}(2011)}]{berthier2011theoretical}%
  \BibitemOpen
  \bibfield  {author} {\bibinfo {author} {\bibfnamefont {L.}~\bibnamefont {Berthier}}\ and\ \bibinfo {author} {\bibfnamefont {G.}~\bibnamefont {Biroli}},\ }\bibfield  {title} {\bibinfo {title} {Theoretical perspective on the glass transition and amorphous materials},\ }\href@noop {} {\bibfield  {journal} {\bibinfo  {journal} {Rev. Mod. Phys.}\ }\textbf {\bibinfo {volume} {83}},\ \bibinfo {pages} {587} (\bibinfo {year} {2011})}\BibitemShut {NoStop}%
\bibitem [{\citenamefont {Pan}\ \emph {et~al.}(2017)\citenamefont {Pan}, \citenamefont {Wu}, \citenamefont {Wang}, \citenamefont {Li},\ and\ \citenamefont {Xu}}]{pan2017structural}%
  \BibitemOpen
  \bibfield  {author} {\bibinfo {author} {\bibfnamefont {S.}~\bibnamefont {Pan}}, \bibinfo {author} {\bibfnamefont {Z.}~\bibnamefont {Wu}}, \bibinfo {author} {\bibfnamefont {W.}~\bibnamefont {Wang}}, \bibinfo {author} {\bibfnamefont {M.}~\bibnamefont {Li}},\ and\ \bibinfo {author} {\bibfnamefont {L.}~\bibnamefont {Xu}},\ }\bibfield  {title} {\bibinfo {title} {Structural origin of fractional stokes-einstein relation in glass-forming liquids},\ }\href@noop {} {\bibfield  {journal} {\bibinfo  {journal} {Sci. Rep.}\ }\textbf {\bibinfo {volume} {7}},\ \bibinfo {pages} {39938} (\bibinfo {year} {2017})}\BibitemShut {NoStop}%
\bibitem [{\citenamefont {Kob}\ and\ \citenamefont {Andersen}(1995{\natexlab{b}})}]{Kob1995Testing2}%
  \BibitemOpen
  \bibfield  {author} {\bibinfo {author} {\bibfnamefont {W.}~\bibnamefont {Kob}}\ and\ \bibinfo {author} {\bibfnamefont {H.~C.}\ \bibnamefont {Andersen}},\ }\bibfield  {title} {\bibinfo {title} {Testing mode-coupling theory for a supercooled binary lennard-jones mixture. {II}. intermediate scattering function and dynamic susceptibility},\ }\href@noop {} {\bibfield  {journal} {\bibinfo  {journal} {Phys. Rev. E}\ }\textbf {\bibinfo {volume} {52}},\ \bibinfo {pages} {4134} (\bibinfo {year} {1995}{\natexlab{b}})}\BibitemShut {NoStop}%
\bibitem [{\citenamefont {Horbach}\ and\ \citenamefont {Kob}(2001)}]{horbach2001relaxation}%
  \BibitemOpen
  \bibfield  {author} {\bibinfo {author} {\bibfnamefont {J.}~\bibnamefont {Horbach}}\ and\ \bibinfo {author} {\bibfnamefont {W.}~\bibnamefont {Kob}},\ }\bibfield  {title} {\bibinfo {title} {Relaxation dynamics of a viscous silica melt: The intermediate scattering functions},\ }\href@noop {} {\bibfield  {journal} {\bibinfo  {journal} {Phys. Rev. E}\ }\textbf {\bibinfo {volume} {64}},\ \bibinfo {pages} {041503} (\bibinfo {year} {2001})}\BibitemShut {NoStop}%
\end{thebibliography}%

\end{document}